\documentclass[%
 reprint,
 amsmath,amssymb,
 aps,
]{revtex4-2}

\usepackage[caption=false]{subfig}

\usepackage{graphicx}
\usepackage[dvipsnames]{xcolor}
\usepackage{dcolumn}
\usepackage{bm}
\usepackage{subfig}
\newcommand{\wBD}{\omega_{\textrm{BD}}}
\newcommand{\rhovac}{\rho_{\textrm{vac}}}

\newcommand{\CC}{\Lambda}
\newcommand{\rv}{\rho_{\rm vac}}
\newcommand{\rvo}{\rho_{\rm vac}^0}
\newcommand{\mpl}{m_{\rm Pl}}
\newcommand{\joan}[1]{{\textcolor{black}{#1}}}
\newcommand{\jtext}[1]{{\textcolor{black}{#1}}}
\newcommand{\jnew}[1]{{\textcolor{black}{#1}}}
\newcommand{\newnew}[1]{{\textcolor{black}{#1}}}

\begin{document}

\preprint{APS/123-QED}

\title{Running vacuum  in Brans \& Dicke theory: a possible cure for the $\sigma_8$ and $H_0$ tensions}
\author{Javier de Cruz P\'erez${}^{1,2,3}$ and Joan Sol\`a Peracaula${}^{2}$ }
\affiliation{
         ${}^{1}$Department of Physics, Kansas State University, 116 Cardwell Hall, Manhattan, KS 66506, USA \\
         ${}^{2}$Departament de F\'isica Qu\`antica i Astrof\'isica, and Institute of Cosmos Sciences, Universitat de Barcelona, Av. Diagonal 647 \\
         ${}^{3}$Departamento de F\' \i sica Te\'orica\\
Universidad Complutense de Madrid, 28040 Madrid, Spain}
\email{decruz@phys.ksu.edu, sola@fqa.ub.edu}

\begin{abstract}
Extensions of the gravitational framework of Brans \& Dicke (BD) are studied by considering two different scenarios: i)  `BD-$\Lambda$CDM',  in which a rigid cosmological constant, $\Lambda$, is included, thus constituting a BD version of the  vanilla concordance $\Lambda$CDM model (the current standard model of cosmology with flat three-dimensional geometry), and ii) `BD-RVM', a generalization of i) in which the vacuum energy density (VED), $\rho_{\textrm{vac}}$, is a running quantity evolving with the square of the Hubble rate:  $\delta\rho_{\textrm{vac}}(H)\propto \nu\, m^2_{\textrm{Pl}} (H^2-H_0^2)$ (with $|\nu|\ll 1$ \textcolor{black}{and $H_0$ being the present value of the Hubble rate}). This dynamical scenario is motivated by recent studies of quantum field theory (QFT) in curved spacetime, which lead to the running vacuum model (RVM).  In both cases, rigid or running $\rho_{\textrm{vac}}$, the GR limit can be recovered smoothly. We solve the background as well as the perturbation equations for each cosmological model and test their performance against the modern wealth of cosmological data, namely a compilation of the latest SNIa+$H(z)$+BAO+LSS+CMB observations. We utilize the AIC and DIC statistical information criteria in order to determine if they can fit better the observations than the concordance model. The two BD extensions are tested by considering three different datasets. According to the AIC and DIC criteria, both BD extensions i) and ii) are competitive, but the second one (the BD-RVM scenario) is particularly favored when it is compared with the vanilla model. This fact may indicate that the current observations favor a mild dynamical evolution of the Newtonian coupling $G_N$ as well as of the VED. 
\jnew{While further studies will be necessary, the results presented here suggest that the Brans \& Dicke theory with running vacuum could have  the potential to alleviate  the two tensions at the same time.}


\end{abstract}

\maketitle


\section{Introduction}\label{sec:introduction}
The so-called concordance model of cosmology, or standard $\CC$CDM model, has been a rather successful  paradigm for the phenomenological description of the universe for more than three decades\,\cite{peebles:1993}, but it became consolidated only in the late nineties\,\cite{Turner:2022gvw}.  Crucial ingredients of it, however,  such as the hypothetical existence  of dark matter (DM),  still lack of direct observational evidence. No less worrisome is the theoretical status of  the cosmological constant (CC) term, $\CC$,  in Einstein's equations. Despite it is sustained by general covariance, nothing is firmly known about its possible physical origin.  The roots of the problem most likely reside in its interpretation  as a quantity which is connected with the vacuum energy density (VED), $\rv$, a fundamental concept in  Quantum Field Theory (QFT).  In fact,  for more than half of a century, the  notion of vacuum energy in cosmology has been a most subtle concept which  has challenged theoretical physicists and cosmologists, especially with the advent of Quantum Theory and in general with the development of the more sophisticated conceptual machinery of  QFT.  The proposed connection between the two quantities is well known: $\rv=\CC/(8\pi G_N)$,  where $G_N$ is Newton's coupling, usually assumed to be constant.  Consistent measurements of $\CC$ (treating it as a fit parameter) made independently in the last twenty years using distant type Ia supernovae (SNIa) and the anisotropies of the cosmic microwave background (CMB), have put the foundations of the concordance  $\CC$CDM model of cosmology\,\cite{Riess:1998cb,Perlmutter:1998np,Boomerang:2000efg,SDSS:2006lmn,WMAP:2008lyn,Riess:2011yx,Planck:2015fie,Planck:2018vyg}.

The above mentioned cosmological observations  have indeed provided strong evidence for a spatially flat and accelerating universe in the present time. The ultimate cause of the acceleration is currently unknown, but it is attributed to an energy component which is popularly  called  ``dark energy" (DE). The DE constitutes $\sim70\%$ of the total energy density of the universe and presumably possessing negative pressure, thereby capable of producing the observed accelerating universe.  The very nature of the DE still remains a complete mystery. The simplest candidate is the aforementioned cosmological term, $\CC$, usually assumed constant and that is why is called the cosmological constant \cite{Peebles:2002gy,Padmanabhan:2002ji}. A flat Friedmann-Lema\^{i}tre-Robertson-Walker (FLRW) model with a cosmological term $(\Lambda$CDM) fits rather accurately the current observational data. However, this model traditionally suffers from the so-called cosmological constant problem\,\cite{Weinberg:1988cp}, which manifests itself in a dual manner: the fine-tuning problem associated to the value of $\CC$ (`the old CC problem'') and what is called the `coincidence problem', see for instance \cite{Sola:2013gha,SolaPeracaula:2022hpd,SolaPeracaula:2023wqw}  for a discussion on the various aspects and implications of the CC problem. In the literature, many other cosmological models have been proposed to describe the DE. In particular, models with time-dependent vacuum energy density seem to be promising, since $\rv(t)$ could have had a high enough value in the very early universe to drive inflation, and subsequently  decay along the expansion history to its small value observed today.  Many of these models, however, are of pure phenomenological nature since the time dependence of $\Lambda(t)$ is parameterized in a totally ad hoc manner and having no obvious connection with any fundamental theory, see e.g. \cite{Bertolami:1986bg,Ozer:1985ws,Peebles:1987ek,Carvalho:1991ut,Espana-Bonet:2003qjh,Wang:2004cp,Borges:2005qs,Sola:2005et,Sola:2005nh,Alcaniz:2005dg,Grande:2006nn,Barrow:2006hia,Costa:2009wv,Grande:2011xf,Bessada:2013maa,Gomez-Valent:2015pia,Rezaei:2019xwo,Macedo:2023zrd} and the old review\,\cite{Overduin:1998zv}. \jnew{In many cases the parameterization is performed through a direct function of the cosmic time or the scale factor, sometimes also as a function of the Hubble parameter, or even some hybrid combination of these possibilities.} 


However, recent calculations involving the renormalization of quantum field theory in FLRW spacetime yield a time-varying $\CC$, and hence a time-varying vacuum as well, in which $\CC$  acquires a dynamical component through quantum effects: $\CC\to \CC+\delta\CC$. The latter evolves through the Hubble rate $H(t)$ as follows: $\delta\Lambda \sim \nu\, (H^2-H_0^2)$  ($|\nu|\ll1$).  This is the characteristic form of the ``running vacuum model'' (RVM). The connection of the latter with QFT can be motivated  from semi-qualitative renormalization group arguments, see  \cite{Sola:2013gha} and references therein. However, an explicit QFT calculation leading to that form has appeared only very recently\,\,\cite{Moreno-Pulido:2020anb,Moreno-Pulido:2022phq,Moreno-Pulido:2022upl,Moreno-Pulido:2023ryo}.  For the latest comprehensive phenomenological analysis of the RVM, see \cite{SolaPeracaula:2023swx}.

Let us also mention a popular class of cosmological scalar field models which were proposed long ago to solve some aspects of the CC problem.  They go under the name of quintessence,
see \cite{Peebles:2002gy} for a review and references.  These models mimic a time-evolving $\CC$ term through a cosmic scalar field and focused essentially on ameliorating the situation with the mentioned cosmic coincidence problem, viz. why $\rvo$ is so close to today's value of the matter density, which falls rapidly with the expansion as $\rho_m\sim a^{-3}$. But again these models are essentially phenomenological in nature since we do not have a viable scalar field candidate to play the role of quintessence in a fundamental context.

All in all, such attempts in different directions have shown that the dynamical DE models  may offer an interesting possibility not only to alleviate some aspects of the cosmological constant and coincidence problems, but also to modulate the processes of structure formation.
A note of caution is in order here, though. Among the above proposals, the very large family of time-varying `$\CC(t)-$models'' should not be confused with the restricted class of the RVMs mentioned above, in which the dynamical dependence  of the VED is derived from the computation and proper renormalization of quantum effects in QFT in curved spacetime. As mentioned, these developments are actually quite recent\,\cite{Moreno-Pulido:2022phq,Moreno-Pulido:2020anb,Moreno-Pulido:2022upl,Moreno-Pulido:2023ryo}.  Therefore, despite some existing  confusion in the literature, the RVMs are to be understood in that (much more) restricted sense, hence closer to fundamental theory,  see \,\cite{Sola:2013gha} for an introduction and \cite{SolaPeracaula:2022hpd,SolaPeracaula:2023wqw} for a review of the recent developments.  Let us also note that, apart from the mentioned QFT framework, a `stringy' version of the RVMs is also available\,\cite{Basilakos:2019acj,Basilakos:2020qmu,Mavromatos:2020kzj,Mavromatos:2021urx}, \newnew{having also potential implications on the cosmological tensions}\,\cite{Gomez-Valent:2023hov,Mavromatos:2023jsb}. The potential dynamics of the cosmic vacuum is therefore well motivated from different fundamental perspectives, and this fact further enhances the interest for the current study.

\indent In this work, indeed, we combine the modern viewpoint of the RVM to revisit  the old  Brans \&  Dicke (BD) framework\,\cite{BransDicke1961} with the purpose  of exploring the consequences of having an evolving gravitational coupling and at the same time a dynamical vacuum.  The BD theory is the simplest extension/modification of Einstein's general theory of relativity (GR). In fact it is the  prototype of scalar-tensor theory of gravitation based on Mach's principle and Dirac's large number hypothesis. Besides, the BD field $\phi$ is not of quantum origin. Rather it is classical in nature and hence can be expected to serve as a relevant candidate to play some role in the late-time evolution of the universe. In this theory, the gravitational constant $G$ is associated with the BD field as $\phi(t) \sim 1/G(t)$ and is coupled to curvature. There is also a new coupling parameter $\wBD$ in front of the  BD kinetic term. It is generally assumed that GR is recovered in the limit $\wBD\rightarrow \infty$, but we actually need also that $\phi(t)$ provides the (inverse) measured value of $G_N$ at present  ($t=t_0$). BD theory is in fact the first historical attempt to extend GR to accommodate a dynamical cosmic variable in the Newtonian coupling $G$, but at that time no cosmological term was included in the study. The interest on this kind of theory was renewed later on, in part owing to its association with superstrings theories, extra-dimensional theories and models with inflation or accelerated expansion \cite{Uehara:1981nq,Freund:1982pg,Banerjee:2000gt,Kim:2004is,Clifton:2005aj,Montenegro:2006yc,Banerjee:2007zd,Olivares:2007rt,Sheykhi:2009yva,Singh:2012zzd,Kumar:2016los,Singh:2016xmc,Ghaffari:2018wks,Mukherjee:2019see}. In recent years, different studies \cite{SolaPeracaula:2018dsw,deCruzPerez:2018cjx,SolaPeracaula:2019zsl,Singh:2019uwv,Singh:2019nok,SolaPeracaula:2020vpg,Kaur:2021dix,Singh:2021jrp} have incorporated the CC term in BD theory as a possibility for explaining the DE phenomena. It remains, of course, to see if realistic mechanisms can be devised that are capable of screening the modified gravity effects at local (astrophysical) scales where deviations from GR are are highly constrained\,see e.g. \cite{Amendola:2015ksp,Clifton:2011jh,Avilez:2013dxa,Gomez-Valent:2021joz} and references therein. We concentrate here only on the cosmological consequences of BD-like theories of gravity.

\indent Specifically, in this work we further dwell upon the BD theory with a cosmological term and investigate in detail two basic cosmological scenarios with flat  FLRW spacetime: i) First we reconsider the BD theory with a rigid CC, $\CC$, along the lines of previous studies such as \cite{SolaPeracaula:2020vpg,SolaPeracaula:2019zsl} and \cite{Singh:2021jrp}.  This will serve as a fiducial BD scenario  maximally close to the $\CC$CDM. It obviously represents a BD generalization of it, which is why we call it  BD-$\CC$CDM; ii) The second scenario adds up more dynamics to the first one along the lines of the RVM, that is to say, we admit that the VED evolves with the cosmic expansion  as $\delta\rho_{\textrm{vac}}\propto \nu\, \mpl^2 (H^2-H_0^2)$ (with $|\nu|\ll1$), just as found in the QFT framework of \cite{Moreno-Pulido:2022phq,Moreno-Pulido:2020anb,Moreno-Pulido:2022upl}. In this way we can explore the BD theory (being the canonical scalar-tensor framework possessing a mild time-evolving gravitational coupling)  by further equipping it with a smooth dynamical vacuum component\,\footnote{We note that this analysis is different from that presented in in \cite{Singh:2021jrp} based on a dynamical $\CC$-term in BD-theory, in which the dependence on $H$ was linear rather than quadratic. The linear model turned out to be not particularly favored.}. We perform the joint statistical analysis using the latest observational data including Type Ia Supernova data, Hubble data, baryon acoustic oscillation data, large scale structure (LSS) formation data and cosmic microwave background data. 

At this point a very much practical motivation for studying the BD-RVM framework must be stood up. Let us recall that apart from the aforementioned theoretical aspects of the cosmological constant problem, it has long been known that  there appear to exist potentially  serious discrepancies  or ``tensions'' between the CMB observations from the Planck Collaboration, which are based on the vanilla $\Lambda$CDM,  and the local direct (distance ladder) measurements of the Hubble parameter today \cite{DiValentino:2020zio}. This brings about the so called $H_0$ tension, which leads to a severe discrepancy at $\sim 5 \sigma$ c.l. Among many possibilities debated in the literature, it has been suggested e.g. that a possible intrinsic evolutionary behavior
of $H_0(z)$ could be underlying this tension\,\cite{Dainotti:2021pqg}, in which case it would not be a discrepancy between local and CMB data, but an effect potentially observable at any
redshift. \textcolor{black}{See the references \cite{Jacques:2013xr,DiValentino:2015sam,Karwal:2016vyq,Bernal:2016gxb,Hart:2017ndk,Jedamzik:2020krr,Gomez-Valent:2020mqn,Carneiro:2018xwq,Banihashemi:2018oxo} and also \cite{DiValentino:2020zio,DiValentino:2021izs} for more possible solutions to the $H_0$ tension.}  In addition, there exist a smaller but appreciable and persistent ($\sim 2-3\, \sigma$) tension in the realm of the large scale structure (LSS) growth data, the $\sigma_8$ tension, \jnew{namely the discrepancy observed in the mean square of matter fluctuations on spheres of radius $R_8 = 8{h^{-1}}$ Mpc}\,\cite{DiValentino:2020vvd}. Specifically, it is concerned with the measurements of weak gravitational lensing at low redshifts ($z<1$). Such a tension is gauged through the parameter $S_8$ or, alternatively, by means of $\sigma_8$; recall that $S_8\equiv\sigma_8\sqrt{\Omega^0_m/0.3}$ (\jnew{$\Omega^0_m$ being the current value of the cosmological matter parameter}). It turns out that these measurements favor weaker matter clustering than that expected from the concordance cosmological model  using parameters determined by CMB measurements. For detailed reviews on these tensions and other challenges afflicting the concordance $\CC$CDM model, \textcolor{black}{ see e.g. ~\cite{DiValentino:2015ola,DiValentino:2015bja, DiValentino:2020vvd, Perivolaropoulos:2021jda,Abdalla:2022yfr,Dainotti:2023yrk,Vagnozzi:2019ezj,Vagnozzi:2023nrq} and the long list of references therein. }

It has been known for quite some time in the literature  that the RVM-type of cosmological models can help in improving the overall fit to the cosmological observations and also in  smoothing out these tensions as compared to the $\CC$CDM. Within the GR-RVM (viz. running vacuum in GR) it  was investigated for instance in \cite{SolaPeracaula:2021gxi,Sola:2017znb,Gomez-Valent:2017idt,Gomez-Valent:2018nib,Sola:2016ecz,Sola:2015wwa,sola2017first,Sola:2017jbl}, see \cite{SolaPeracaula:2022mlg,SolaPeracaula:2018xsi} for a summary. When we move to BD theory with a cosmological constant, i.e. the BD-$\CC$CDM, the model turns out to mimic the RVM and improves the fits as well\,\cite{Sola:2021txs,SolaPeracaula:2020vpg,SolaPeracaula:2019zsl}. It is therefore natural to keep on exploring the RVM framework and perform the next step.  Thus, in the present paper we address for the first time the running vacuum in the BD context, i.e. the BD-RVM scenario,  and check if further improvement (relieving)  of the tensions can be attained as compared to the BD-$\CC$CDM. We find indeed a significant improvement of the current status of the mentioned tensions.

\indent The paper is organized as follows. In Sec. \ref{sec:cosmological_equations}, we present the models and the corresponding field equations within the context of Brans \& Dicke theory. Section \ref{sec:effective_EoS} is devoted to discuss the effective equation of state (EoS) of the extended BD models.  Being the large scale structure (LSS) growing data indispensable for a proper discussion of the $\sigma_8$ tension, in Sec. \ref{sec:structure_formation} we describe our treatment of the matter perturbations. In Section \ref{sec:Data_and_methodology}  we enumerate and briefly describe the different sources of observational data used in this paper and the method employed to constrain the free parameters of the models. At the same time we define the three datasets that will be used throughout our comparative model analysis.  The results are discussed in detail in Section \ref{sec:discussion}. In Section \ref{sec:conclusions}  we summarize our findings and present the main conclusions of this work. For completeness, in the appendix we summarize the numerical results for the case when we include the polarization data in the CMB analysis.
\vspace{0.5cm}
\section{BD field equations with dynamical vacuum }\label{sec:cosmological_equations}
We consider a spatially flat  homogeneous and isotropic universe described by the  FLRW metric
\begin{equation}\label{eq:FLRW_metric}
ds^2 = -dt^2 + a^2(t) \left[dr^2+r^2(d\theta^2+\textrm{sin}^2\theta d\phi^2)\right],
\end{equation}
where $(r,\theta, \phi)$ are defined as the comoving coordinates, $t$ denotes the cosmic time and $a(t)$ is the scale factor of the universe. Throughout this work we consider the natural units convention, namely: $\hbar = c=1$.
The gravity theory of Brans \& Dicke \cite{BransDicke1961}, contains an additional {\it d.o.f.} with respect to General Relativity, which is encoded in the BD-field denoted by $\phi$. The BD-action, written in the Jordan frame and in the presence of the vacuum energy density $\rhovac$ takes the following form:
\begin{equation}\label{BD_action}
\begin{split}
S_{\rm BD}=\int d^4x \sqrt{-g}\Big[\frac{1}{16\pi}\left(\phi R-\frac{\wBD}{\phi}g^{\mu\nu}\partial_\mu\phi\partial_\nu\phi \right)\\
-\rv +\mathcal{L}_{m}\Big]\,.
\end{split}
\end{equation}
%
The scalar field, which is minimally coupled to the curvature, has dimension 2 in natural units and it represents, at any time, the inverse value of the Newtonian gravitational coupling $\phi(t) = 1/G(t)$. The dimensionless parameter in front of the kinetic term, $\wBD$, is the BD-parameter and $\mathcal{L}_m$ contains the contribution from the matter fields.  

Variation of the action \eqref{BD_action} with respect to the metric $g_{\mu\nu}$ and the BD-scalar field $\phi$, respectively, yields the  field equations:
\begin{equation}\label{eq:BD_field_equations}
G_{\mu\nu}\equiv R_{\mu\nu}-\frac{1}{2} g_{\mu\nu}R =\frac{8\pi}{\phi}\left(\tilde{T}_{\mu\nu}+T^{\textrm{BD}}_{\mu\nu}\right),
\end{equation}
and
\begin{equation}\label{eq:KG_equation}
\Box\phi=\frac{8\pi}{(2\wBD+3)}\tilde{T}.
\end{equation}
In the above equations,
\begin{equation}\label{eq:EM_tensor}
\tilde{T}_{\mu\nu} = p{g_{\mu\nu}} + (\rho+p)U_{\mu}U_{\nu}    
\end{equation}
is the total energy-momentum tensor, with $\tilde{T}\equiv \tilde{T}^{\mu}_{\mu}$ its trace. In it,  $\rho\equiv \rho_m + \rho_r + \rhovac$ and $p\equiv p_m + p_r + p_{\textrm{vac}}$ represent the total (proper) density and pressure, whose matter and vacuum components satisfy the usual equations of state  $p_m = 0 $,  $p_r = (1/3)\rho_r$ and $p_{\textrm{vac}} = -\rhovac$. The term $\rho_m$ involves the contribution from cold dark matter (CDM) and baryons,  whereas $\rho_r$ is the relativistic component from photons and massless neutrinos.  Furthermore, in Eq.\eqref{eq:BD_field_equations}, the piece $ T^{\textrm{BD}}_{\mu\nu}$ stands for the remaining term that emerges from the metric functional variation of the action \eqref{BD_action} and its structure depends entirely on $\phi$ and its derivatives. Such a piece defines the energy-momentum tensor of the BD-field:
\begin{widetext}
\begin{equation}
T^{\textrm{BD}}_{\mu\nu}= \frac{1}{8\pi}\left[\frac{\wBD}{\phi}\left(\nabla_{\mu}\phi\nabla_{\nu}\phi-\frac{1}{2}g_{\mu\nu}\nabla_{\alpha}\phi
\nabla^{\alpha}\phi\right)
 + \left(\nabla_{\mu}\nabla_{\nu}\phi-g_{\mu\nu}\nabla_{\alpha}\nabla^{\alpha}\phi \right)\right].
\end{equation}
\end{widetext}
For $\phi=$const. the above energy-momentum tensor vanishes, and when such a constant is precisely $\phi=1/G_N$, Eq.\,\eqref{eq:BD_field_equations} boils down to the ordinary Einstein's field equations, i.e. we recover GR.  Here $G_N=1/m^{2}_{\textrm{Pl}}$  is the measured value of Newton's constant, with $m_{\textrm{Pl}}\simeq 1.2\times 10^{19}$ GeV the ordinary Planck mass.  Notice, however,  that on implementing the GR limit it is mandatory to set also $\wBD\to\infty$ such that the wave equation for the BD-field  \eqref{eq:KG_equation} consistently decouples at the same time. For variable $\phi(t)$, we must recover too the measured gravity value $1/\phi(t_0)=G(t_0)=G_N$ at the present cosmic time $t_0$, of course, but $G(t)\equiv 1/\phi(t)$  can be slightly different from $G_N$ at other times. Such small difference, however, can impinge positively on our analysis.

In the framework of the FLRW metric \eqref{eq:FLRW_metric}, the cosmological equations that govern the dynamics of the universe within BD gravity can be written as follows:
\begin{equation}\label{eq:Friedmann_equation}
3H^2+3H\frac{\dot{\phi}}{\phi}-\frac{\wBD}{2}\frac{\dot{\phi^2}}{\phi^2}
=\frac{8\pi}{\phi}\rho
\end{equation}
\begin{equation}\label{eq:pressure_equation}
2\dot{H}+3H^2+\frac{\ddot{\phi}}{\phi}+2H\frac{\dot{\phi}}{\phi}+\frac{\wBD}{2}\frac{\dot{\phi^2}}{\phi^2}=-\frac{8\pi}{\phi} p
\end{equation}
\begin{equation}\label{eq:SF_equation}
\ddot{\phi}+3H\dot{\phi}=
\frac{8\pi}{(2\wBD+3)}(\rho-3p)
\end{equation}
where the overdots denote derivatives with respect to the cosmic time $t$ and $H=\dot{a}/a$ is the Hubble parameter. Once more we can check that when we set $\phi=1/G_N$ {\it and} at the same time $\wBD\rightarrow\infty$,  we do in fact retrieve the GR limit, which in this case is obtained within the FLRW metric. Indeed, in the mentioned limit we find that equations \eqref{eq:Friedmann_equation} and \eqref{eq:pressure_equation} render the standard Friedmann and pressure equations respectively, whereas \eqref{eq:SF_equation} consistently decouples. \\

No interaction between the BD-field and the ordinary matter and vacuum is considered,  so from the Bianchi identity we get $\nabla^{\mu}\tilde{T}_{\mu\nu} = 0$. This can be explicitly verified with the above field equations.  We assume that baryons and radiation (photons and relativistic neutrinos) are covariantly conserved separately, which leads to the usual local energy conservation equations, namely: 
\begin{equation}\label{eq:phoronconservation}
\dot{\rho}_b +3H\rho_b=0   
\end{equation} 
and
\begin{equation}\label{eq:baryonconservation}
\dot{\rho}_r +4H\rho_r=0\,.   
\end{equation}
It should be mention at this point that despite neutrinos do
not behave as pure radiation in the late universe, the
ratio between the energy density of non-relativistic neutrinos to the total energy density of non-relativistic matter (baryons plus CDM) remains very small throughout the entire cosmic
history up to our time  (below $10^{-3}$)\,\cite{SolaPeracaula:2021gxi}, and therefore for our purposes in this paper we can safely treat neutrinos as if being relativistic all the time.

In contradistinction to baryons and radiation, cold dark matter does not need to be conserved in general and  it may interact with vacuum, similarly as in the analysis of the GR-RVM performed in \cite{Sola:2017jbl}.  So for CDM we have,  in general,  the following modified conservation equation:
\begin{equation}\label{eq:cdm_and_vacuum_conservation_equation}
\dot{\rho}_{\textrm{cdm}}+3H\rho_{\textrm{cdm}} = -\dot{\rho}_{\textrm{vac}}\,.   
\end{equation}
Notice that in the case of the BD-$\CC$CDM the \textit{r.h.s.} of Eq.\,\eqref{eq:cdm_and_vacuum_conservation_equation} is zero and  the CDM is conserved. Not so, of course,  when $\rho_{\textrm{vac}}$ changes with the cosmological expansion, as e.g. in the BD-RVM context.   Both options will be considered separately in the next sections.

Unfortunately it is not possible to obtain analytical solutions for the system of differential equations \eqref{eq:Friedmann_equation}-\eqref{eq:pressure_equation}-\eqref{eq:SF_equation}. However, it is reasonable to assume that the evolution of the BD-field $\phi(a)$ should be close around to the present value $\phi(a=1)=1/G_N$, as otherwise the departures of $1/\phi(a)$ from $G_N$ would be significant and would have been detected experimentally. Bearing this in mind, it is meaningful to search for solutions of the field equations in power-law form.  We proceed as in \cite{SolaPeracaula:2018dsw,deCruzPerez:2018cjx} with the following ansatz:
\begin{equation}\label{eq:power-law_BD_field}
\phi(a) = \phi_0{a^{-\epsilon}} = \frac{1}{G_N}a^{-\epsilon} \quad (|\epsilon|\ll 1)\,,
\end{equation}
where $\epsilon$ is a constant whose value should be small in order to be consistent with the measurements of the Newtonian coupling.  For $\epsilon>0$, $\phi$ decreases (and hence Newton's coupling increases) with the expansion and this means that it has an asymptotically free behavior in the past (where it lies the most energetic epoch of the universe) since it becomes smaller. For $\epsilon<0$, the opposite occurs, that is to say, in such case the gravitational coupling decreases with the expansion.

If we insert \eqref{eq:power-law_BD_field} into the set of equations \eqref{eq:Friedmann_equation}-\eqref{eq:pressure_equation}-\eqref{eq:SF_equation} we arrive at the equations

\begin{eqnarray}
&H^2 = \frac{8\pi}{3}\frac{1}{\beta\phi}(\rho_m+\rho_r+\rhovac)\\
&\dot{H}(2-\epsilon) + H^2\left(3-2\epsilon + \epsilon^2 +(1/2)\wBD\epsilon^2\right) \\ &= -\frac{8\pi}{\phi}\left(\frac{\rho_r}{3}-\rhovac\right)\nonumber\\
&-\epsilon\dot{H} + (-3\epsilon+\epsilon^2)H^2 = \frac{8\pi}{2\wBD +3}\frac{1}{\phi}(\rho_m + 4\rhovac). 
\end{eqnarray}
We recall that $\rho_m = \rho_b + \rho_{\textrm{cdm}}$. In the above equations we have introduced the following definitions
\begin{equation}\label{eq:beta_and_nueff_definitions}
\beta \equiv 1-\nu_{\textrm{eff}}, \qquad \nu_{\textrm{eff}} \equiv \epsilon\left(1+\frac{\wBD}{6}\epsilon\right).    
\end{equation}
As in \cite{deCruzPerez:2018cjx} we assume that $|\wBD\epsilon|\sim 1$, consequently we have $|\wBD\epsilon|>|\epsilon|\gg |\epsilon^2|$ and $|\wBD\epsilon^2|\sim |\epsilon|$. Taking into account this hierarchy of orders of magnitudes, it is possible to get a relation  between $\wBD$, $\epsilon$ and the current values of the non-relativistic and radiation energy densities, see \cite{deCruzPerez:2018cjx} for the details:
\begin{equation}\label{eq:relation_wBDeps}
\wBD\epsilon = -\frac{12 -9\Omega^0_m-12\Omega^0_r}{6-3\Omega^0_m -4\Omega^0_r} + \mathcal{O}(\epsilon).
\end{equation}
From the above expression it is clear that $|\wBD\epsilon|\sim\mathcal{O}(1)$. The final numerical values are to be determined when the cosmological models under study are tested against the observational data considered. 
Hereafter we consider two different cosmological models both in the theoretical framework of the Brans \& Dicke modified gravity. Additionally, we also provide the results for the standard model of cosmology which in this paper is denoted by GR-$\Lambda$CDM model. 
\subsection{BD-$\Lambda$CDM (constant vacuum)  $\rhovac(a) = \rho^0_{\textrm{vac}}=\textrm{const.}$}
In this scenario, vacuum is not in interaction with cold dark matter and the value of its energy density remains constant. It is important to remark that in the original formulation of Brans \& Dicke \cite{BransDicke1961} the cosmological term was not included so we may consider the BD-$\Lambda$CDM model as the minimal extension of the BD model in order to predict the observed accelerated expansion of the universe. See \cite{SolaPeracaula:2019zsl} and \cite{SolaPeracaula:2020vpg} for a detailed study of the BD-$\Lambda$CDM model. Considering the usual definitions for the mass parameter of each one of the species at present time $\Omega^0_i = \rho^0_i/\rho^0_c$, where $i$ runs for $m,\textrm{vac},r$ and $\rho^0_c=3H^2_0/8\pi{G_N}$ is the current critical energy density we can express the normalized Hubble function $E(a)\equiv H(a)/H_0$ as:
\begin{equation}\label{eq:E_function_BD-LCDM}
E^2(a) = \frac{a^{\epsilon}}{\beta}\left[\Omega^0_m{a^{-3}} + \Omega^0_r{a^{-4}} + \Omega^0_{\textrm{vac}} \right].    
\end{equation}
As expected the cosmic sum rule for $\Omega^0_i$ does not have the usual form due to the presence of $\epsilon$ and $\wBD$
\begin{equation}\label{eq:sum_rule_BD-LCDM}
\Omega^0_m+\Omega^0_r+\Omega^0_{\textrm{vac}} = \beta = 1-\nu_{\textrm{eff}}.     
\end{equation}
The GR-$\Lambda$CDM limit, $E^2(a) = \Omega^0_m{a^{-3}} + \Omega^0_r{a^{-4}} + \Omega^0_{\textrm{vac}}$, is recovered by setting $\epsilon\rightarrow 0$ and $\beta\rightarrow 1$. 
\subsection{BD-RVM (running vacuum) $\rhovac(H) = \frac{3\phi}{8\pi}\left(c_0 + \nu{H^2}\right)$}
In this alternative scenario the cosmological term is no longer constant.  It is nevertheless assumed to be essentially constant up to a dynamical component which evolves as the square of the Hubble rate:
\begin{equation}\label{eq:LambdaRVM}
\Lambda(H) = 3c_0 + 3\nu{H^2}\,,
\end{equation}
where the coefficient $3$ is for convenience. Thus, the vacuum energy density is  $\rv(H)=\CC(H)\phi/(8\pi)$ and since $\phi=1/G\sim \mpl^2$ it evolves as $\delta\rho_{\textrm{vac}}\sim 3\,\nu\, \mpl^2 (H^2-H_0^2)$ when we compare its value at $H$ with its current value at $H_0$. This form is characteristic of the RVM, as discussed previously. The dimensionless parameter $\nu$ must be fixed by comparing with observations and we expect it to have a small value ($|\nu|\ll1$), typically $\nu\sim 10^{-3}$. This value is compatible with recent BBN bounds on modified gravity models\,\cite{Asimakis:2021yct}.  Because the vacuum is slightly time-dependent now, we shall assume that it is in interaction with cold dark matter (CDM) through the generalized local conservation law \eqref{eq:cdm_and_vacuum_conservation_equation}, where now we have the specific RVM  form for $\rho_{\rm vac}$. We can fix the value of the constant $c_0$  (of dimension 2 in natural units) by imposing the boundary condition $\rhovac(a=1)=\rhovac^0$ which leads us to the following equality $c_0=H^2_0(\Omega^0_{\textrm{vac}}-\nu)$, where $\Omega^0_{\textrm{vac}}$ is $\rvo$ in units of the current critical density $\rho^0_c$. Notice that the vacuum in Eq.\eqref{eq:LambdaRVM} is dominated by the constant term $c_0$, whilst $\nu H^2$ is a subdominant component.  The fact of having $c_0\neq 0 $ is essential in order to remain close to the concordance model and hence be able to fit the large scale structure observational data comparatively well. The dimensionless coefficient $\nu$ can be interpreted as the $\beta$-function of the running of the vacuum at low energies and its theoretical value can be estimated in the context of QFT in curved space times see \cite{Moreno-Pulido:2020anb,Moreno-Pulido:2022phq,Moreno-Pulido:2022upl} and references therein. However, here we determine its value phenomenologically by testing the theoretical predictions of the BD-RVM cosmological model against the latest cosmological data. 

A closed-form formula for the normalized Hubble function in the BD-RVM framework can be explicitly derived  using the equations of the previous section, with the following result:
\begin{equation}\label{eq:Hubble_function_RVM}
\begin{array}{ll}
E^2(a) =&1 + \frac{1}{\beta -\nu}\Big[\Omega^0_m(a^{-C_1 +\epsilon} -1)  + \alpha(1 - a^{-C_1 +\epsilon}) \\
   & + \, \Omega^0_r\left(  \frac{\beta -\nu}{\beta + 3\nu}a^{-4+\epsilon}  + \frac{4\nu}{\beta + 3\nu}a^{-C_1+\epsilon} -1 \right) \Big]\,.
\end{array}
\end{equation}
where to simplify the formulae, we have introduced the following definitions
\begin{equation}
C_1\equiv 3\left(\frac{\beta-\nu}{\beta}\right) \qquad \alpha = -\epsilon\frac{\Omega^0_{\textrm{vac}} - \nu}{C_1-\epsilon}.    
\end{equation}
\\
In this case the GR-$\Lambda$CDM limit is recovered by setting $\epsilon, \nu\rightarrow 0$ and $\beta\rightarrow 1$, which implies $C_1=3$ and $\alpha=0$.

We should also stress that the dynamical character of the vacuum energy density through the square of the Hubble rate $\sim H^2$, as implied by the RVM form \eqref{eq:LambdaRVM}, is highly convenient since it is compatible with the general covariance of the effective action, as demonstrated also from the explicit calculations of\,\cite{Moreno-Pulido:2020anb,Moreno-Pulido:2022phq,Moreno-Pulido:2022upl,Moreno-Pulido:2023ryo}.  In contrast, a time-dependence of the VED through a linear term $\sim H$ is not compatible with general covariance. This does not make a model of this sort a priori, though,  since a linear term can be admitted on phenomenological grounds and may be associated with e.g. dissipative effects produced by bulk viscosity\,\cite{Ren:2006mb,Komatsu:2013qia}.  However, even on phenomenological grounds such linear dependence on $H$ is not favored, see e.g. \cite{Montenegro:2006yc} and especially the recent analysis \cite{Singh:2021jrp}. In point of fact, the difficulties with the linear term $\sim H$ as part of the VED have also been studied at length in the GR context, see\,\cite{Gomez-Valent:2014fda}, where it has been shown that it leads to serious problems with the description of the LSS data. In contrast, as we shall see here, the $\sim H^2$ component in the BD-RVM framework fits extremely well the cosmological observations.  This turns out to be one of the main results of the present work. 

The convenience of the quadratic component in the Hubble rate is in correspondence with the good results obtained for the RVM in the GR context \cite{SolaPeracaula:2021gxi}. The BD framework adds up to it the dynamics of the gravitational coupling, and this feature proves very helpful to further optimize the phenomenological performance of the model. We shall substantiate this assertion in detail throughout the present analysis.

\begin{table*}
\begin{ruledtabular}
\begin{tabular}{cccc}
\\[+0.5mm]
\multicolumn{1}{c}{} & \multicolumn{3}{c}{\Large Baseline}
\\[+0.5mm]
\\\hline
\\[+1mm]
{ \bf Parameter} & { \bf GR-$\Lambda$CDM}  & { \bf BD-$\Lambda\textrm{CDM}$ }   & { \bf BD-$\textrm{RVM}$}\\  
\\[+1mm]
{ $\omega_b$}  & { $0.02213 (0.02213) \pm 0.00020$}  & { $0.02207 (0.02208)\pm 0.00020$}    & { $0.02201 (0.02201)\pm 0.00022$}\\
{$n_s$}   & {$0.9665 (0.9666) \pm 0.0042$} & {$0.9639 (0.9639)\pm 0.0044$} & {$0.9620 (0.9618)\pm 0.0050$}\\
{$10^{9}A_s$}    & {$2.054 (2.054) \pm 0.033$} & {$2.081 (2.081)\pm 0.036$}  & { $2.085 (2.085)\pm 0.036$}\\
{$H_0$[km/s/Mpc]}   & {$68.56 (68.57) \pm 0.41$} & { $66.73 (66.75)\pm 0.81$}  & { $68.0 (68.1)^{+1.4}_{-1.2}$}\\  
{$\Omega^0_m$}   & { $0.3010 (0.3008)\pm 0.0052$} & { $0.3031 (0.3030)\pm 0.0053$}   & { $0.3041 (0.3041)\pm 0.0054$}\\  
{$\sigma_8(0)$}   & {$0.7997 (0.7997)\pm 0.0071$} & { $0.758 (0.758)\pm 0.018$}     & { $0.751 (0.749)\pm 0.019$}\\   
{$S_8$}   & {$0.801 (0.801)\pm 0.011$} & { $0.761 (0.762)\pm 0.019$}     & { $0.768 (0.753)\pm 0.018$}\\   
{$\epsilon$}   & {-} & { $0.0039 (0.0038)\pm 0.0015$}     & { $-0.0016 (-0.0022)^{+0.0037}_{-0.0049}$}\\   
{$\nu$}   & {-} & {-}    & {$0.0019 (0.0021)^{+0.0016}_{-0.0013}$}\\   
{$\chi^2_{\rm min}$}   & { 71.83} & { 64.68}    & { 63.15}\\  
{$\Delta$AIC}   & { -} & { 4.76}     & { 3.81 }\\
{$\Delta$DIC}   & { -} & { 5.06}     & { 5.10 }\\
\end{tabular}
\\[+1mm]
\end{ruledtabular}
\caption{The mean-fit values and the 68\% confidence limits for the characterizing parameters of the cosmological models under study upon consideration of our Baseline dataset, which contains data from the string SNIa+$H(z)$+BAO+LSS+CMB. \textcolor{black}{We also include, in parentheses, the best-fit values}. We provide the values of the parameters common to all the cosmological models, to wit: the baryonic density parameter $\omega_b=\Omega^0_b{h^2}$, the spectral tilt parameter $n_s$ and the amplitude parameter $A_s$ of the primordial power spectrum, the Hubble parameter $H_0$ and the normalized matter density parameter $\Omega^0_m$. Note that we do not consider the optical depth parameter $\tau$ since it is not sensitive to our analysis. In addition to the before mentioned parameters the BD-$\Lambda$CDM model contains an extra {\it d.o.f} contained in $\epsilon$ and the BD-RVM model has two extra parameters, $\epsilon$ and $\nu$. We also display for all the models the value of the derived parameters $\sigma_8(0)$ and $S_8\equiv\sigma_8\sqrt{\Omega^0_m/0.3}$. $\chi^2_{\textrm{min}}$ denotes the minimum value of the $\chi^2$-function and finally $\Delta$AIC and $\Delta$DIC represent the differences of the AIC and DIC values with respect to the GR-$\Lambda$CDM model.} 
\label{tab:table_1}
\end{table*}

\begin{table*}
\begin{ruledtabular}
\begin{tabular}{cccc}
\\[+0.5mm]
\multicolumn{1}{c}{} & \multicolumn{3}{c}{\Large Baseline+$S_8$}
\\[+0.5mm]
\\\hline
\\[+1mm]
{ \bf Parameter} & { \bf GR-$\Lambda$CDM}  & { \bf BD-$\Lambda\textrm{CDM}$ }   & { \bf BD-$\textrm{RVM}$}\\  
\\[+1mm]
{ $\omega_b$}  & { $0.02215 (0.02215)\pm 0.00020$}  & { $0.02207 (0.02208)\pm 0.00020$}    & { $0.02200 (0.02199)\pm 0.00021$}\\
{$n_s$}   & {$0.9670 (0.9670)\pm 0.0042$} & {$0.9639 (0.9639)\pm 0.0044$} & {$0.9613 (0.9614)\pm 0.0047$}\\
{$10^{9}A_s$}    & {$2.046 (2.046)\pm 0.034$} & {$2.080 (2.080)\pm 0.035$}  & { $2.085 (2.084)\pm 0.036$}\\
{$H_0$[km/s/Mpc]}   & {$68.71 (68.70)\pm 0.40$} & { $66.65 (66.65)\pm 0.80$}  & { $68.35 (68.32)\pm 0.71$}\\  
{$\Omega^0_m$}   & { $0.2990 (0.2989)\pm 0.0050$} & { $0.3026 (0.3025)\pm 0.0052$}   & { $0.3044 (0.3042)\pm 0.0054$}\\  
{$\sigma_8(0)$}   & {$0.7970 (0.7971)\pm 0.0071$} & { $0.754 (0.754)\pm 0.016$}     & { $0.751 (0.734)\pm 0.019$}\\   
{$S_8$}   & {$0.796 (0.796)\pm 0.011$} & { $0.757 (0.757)\pm 0.017$}     & { $0.756 (0.739)\pm 0.020$}\\   
{$\epsilon$}   & {-} & { $0.0042 (0.0041)\pm 0.0014$}     & { $-0.0031 (-0.0029) \pm 0.0019$}\\   
{$\nu$}   & {-} & {-}    & {$0.00237 (0.00228)\pm 0.00080$}\\   
{$\chi^2_{\rm min}$}   & { 74.42} & { 65.02}    & { 63.17}\\  
{$\Delta$AIC}   & { -} & { 7.01}     & { 6.39 }\\
{$\Delta$DIC}   & { -} & { 7.61}     & { 6.59 }\\
\end{tabular}
\\[+1mm]
\caption{\label{tab:table_2} As in Table \ref{tab:table_1} but adding as an input the $S_8\equiv\sigma_8\sqrt{\Omega^0_m/0.3} = 0.737^{+0.040}_{-0.036}$ weak-lensing measurement from \cite{Hildebrandt:2018yau}.}
\end{ruledtabular}
\label{tab:withS8}
\end{table*}
\begin{table*}
\begin{ruledtabular}
\begin{tabular}{cccc}
\\[+0.5mm]
\multicolumn{1}{c}{} & \multicolumn{3}{c}{\Large Baseline+$H_0$}
\\[+0.5mm]
\\\hline
\\[+1mm]
{ \bf Parameter} & { \bf GR-$\Lambda$CDM}  & { \bf BD-$\Lambda\textrm{CDM}$ }   & { \bf BD-$\textrm{RVM}$}\\  
\\[+1mm]
{ $\omega_b$}  & { $0.02228 (0.02228)\pm 0.00020$}  & { $0.02228 (0.02228)\pm 0.00020$}    & { $0.02205 (0.02193)\pm 0.00020$}\\
{$n_s$}   & {$0.9693 (0.9693)\pm 0.0042$} & {$0.9692 (0.9692)\pm 0.0043$} & {$0.9619 (0.9585)\pm 0.0045$}\\
{$10^{9}A_s$}    & {$2.058 (2.058)\pm 0.033$} & {$2.060 (2.059)\pm 0.035$}  & { $2.078 (2.080)\pm 0.036$}\\
{$H_0$[km/s/Mpc]}   & {$69.13 (69.13)\pm 0.38$} & { $69.02 (69.06)\pm 0.67$}  & { $70.34 (71.39)\pm 0.63$}\\  
{$\Omega^0_m$}   & { $0.2942 (0.2941)\pm 0.0048$} & { $0.2942 (0.2941)\pm 0.0048$}   & { $0.2994 (0.3021)\pm 0.0050$}\\  
{$\sigma_8(0)$}   & {$0.7970 (0.7971)\pm 0.0072$} & { $0.795 (0.795)\pm 0.016$}     & { $0.764 (0.649)\pm 0.018$}\\  
{$S_8$}   & {$0.789 (0.789)\pm 0.011$} & { $0.757 (0.757)\pm 0.017$}     & { $0.763 (0.648)\pm 0.019$}\\   
{$\epsilon$}   & {-} & { $0.0002 (0.0002)\pm 0.0013$}     & { $-0.00767 (-0.0104)^{+0.00069}_{-0.00220}$}\\   
{$\nu$}   & {-} & {-}    & {$0.00332 (0.00492)^{+0.00084}_{-0.00065}$}\\   
{$\chi^2_{\rm min}$}   & { 87.97} & { 87.95}    & { 70.68}\\  
{$\Delta$AIC}   & { -} & { -2.37}     & { 12.43 }\\
{$\Delta$DIC}   & { -} & { -2.46}     & { 12.10 }\\
\end{tabular}
\\[+1mm]
\caption{\label{tab:table_3} The same as in Table \ref{tab:table_1} but considering the prior on $H_0=73.04\pm 1.04$ km/s/Mpc  from \cite{Riess:2021jrx}.}
\end{ruledtabular}
\label{tab:withHo}
\end{table*}
\begin{figure*}[htbp]
\subfloat[\centering]{{\includegraphics[width=9.5cm, height = 7.5cm  ]{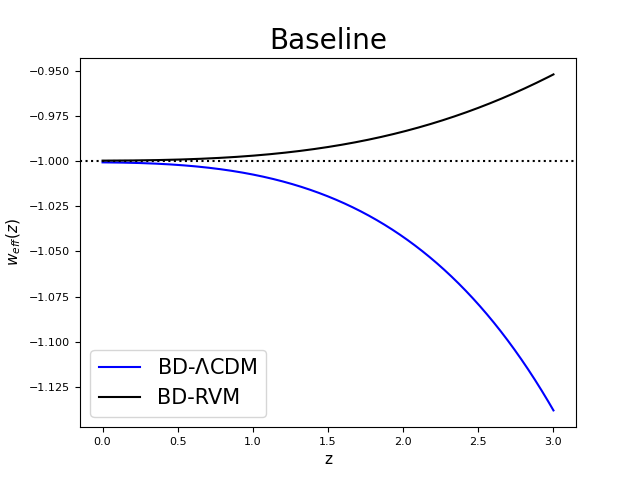} }}
\subfloat[\centering]{{\includegraphics[width=9.5cm, height = 7.5cm ]{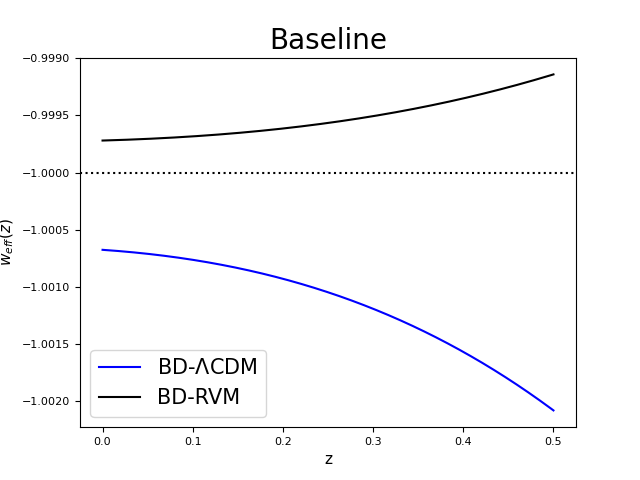} }}
\caption{\label{fig:w_eff_Baseline}{\it (a) plot}: Effective equation of state parameter \eqref{eq:effective_EoS} for the BD-$\Lambda$CDM and the BD-RVM cosmological models considering the results obtained with the Baseline set. 
{\it (b) plot}: Magnifies the region of {\it (a)} near our time.  }    

\end{figure*}
\begin{figure*}[htbp]
\subfloat[\centering]{{\includegraphics[width=9.5cm, height = 7.5cm]{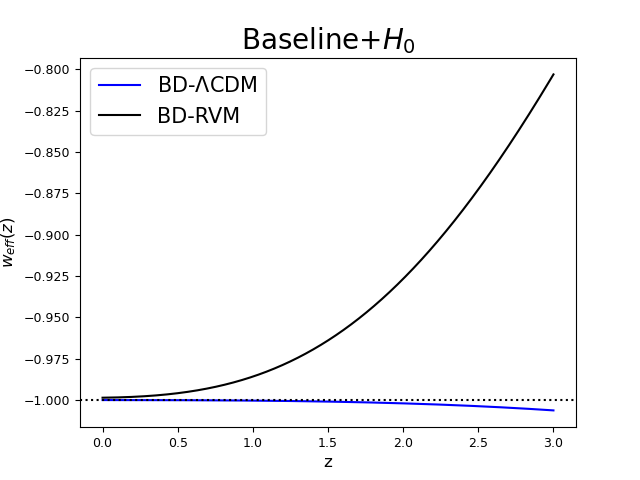} }}
\subfloat[\centering]{{\includegraphics[width=9.5cm, height = 7.5cm]{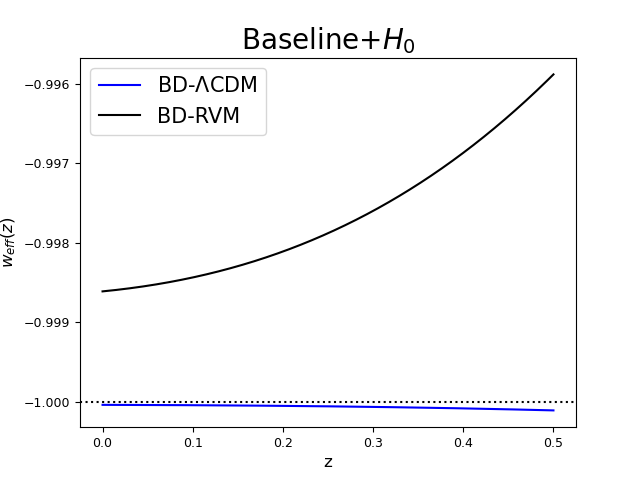} }}
\caption{\label{fig:w_eff_Baseline_H0}The same as in Fig.\,\ref{fig:w_eff_Baseline} but considering the results obtained with the Baseline+$H_0$ set. }
\label{fig:example}
\end{figure*}
\begin{figure*}[htbp]
\centering
\mbox{\includegraphics[width=170mm]{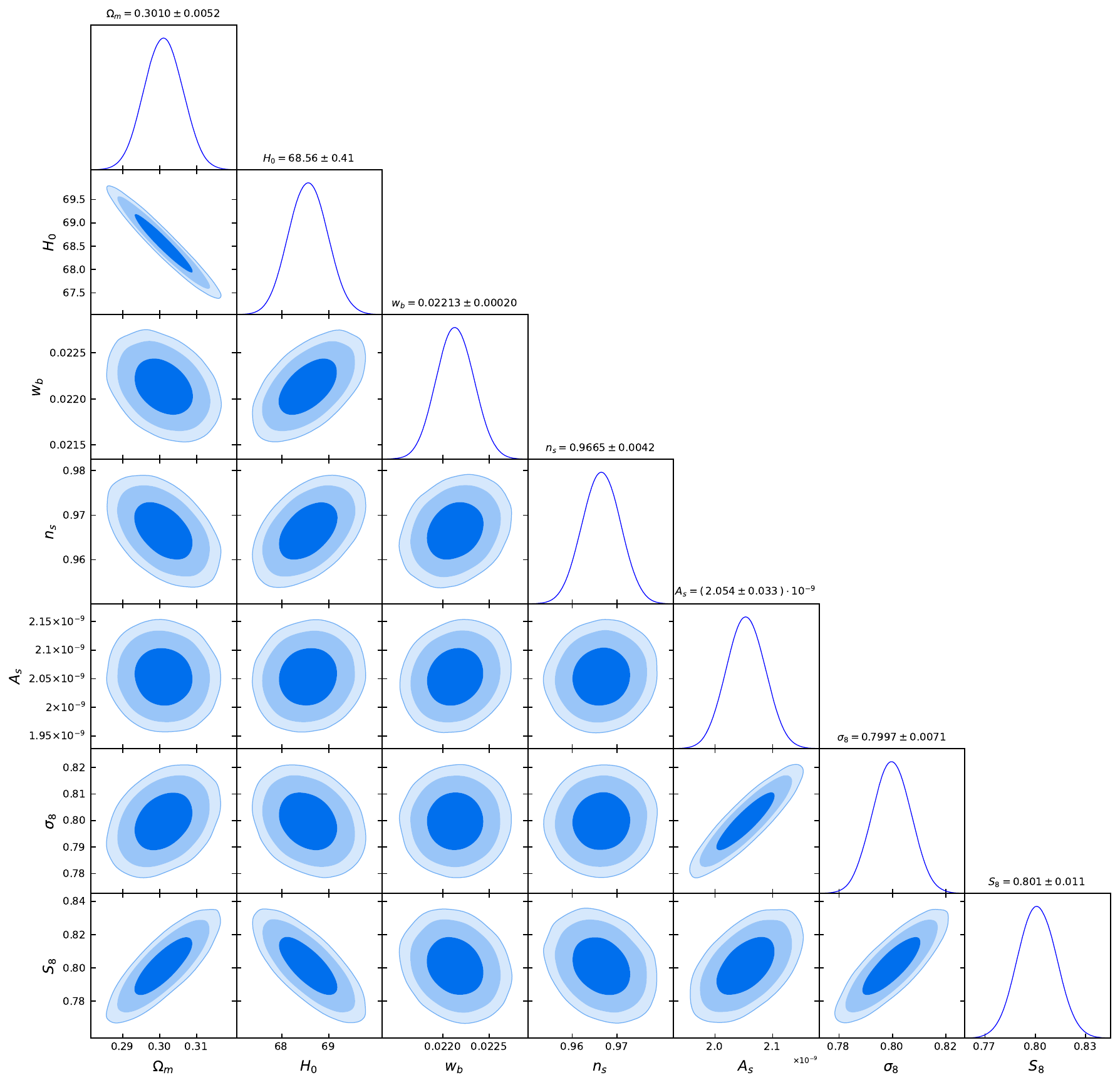}}
\caption{Two-dimensional marginalized likelihood distributions, for each one of the parameters, in the GR-$\Lambda$CDM model at 1$\sigma$, 2$\sigma$ and 3$\sigma$ c.l., obtained considering the Baseline dataset. We also included the corresponding one-dimensional marginalized likelihood distributions. The $H_0$ parameter is expressed in km/s/Mpc. }
\label{fig:GR_LCDM_Baseline}
\end{figure*}
\begin{figure*}[htbp]
\centering
\mbox{\includegraphics[width=170mm]{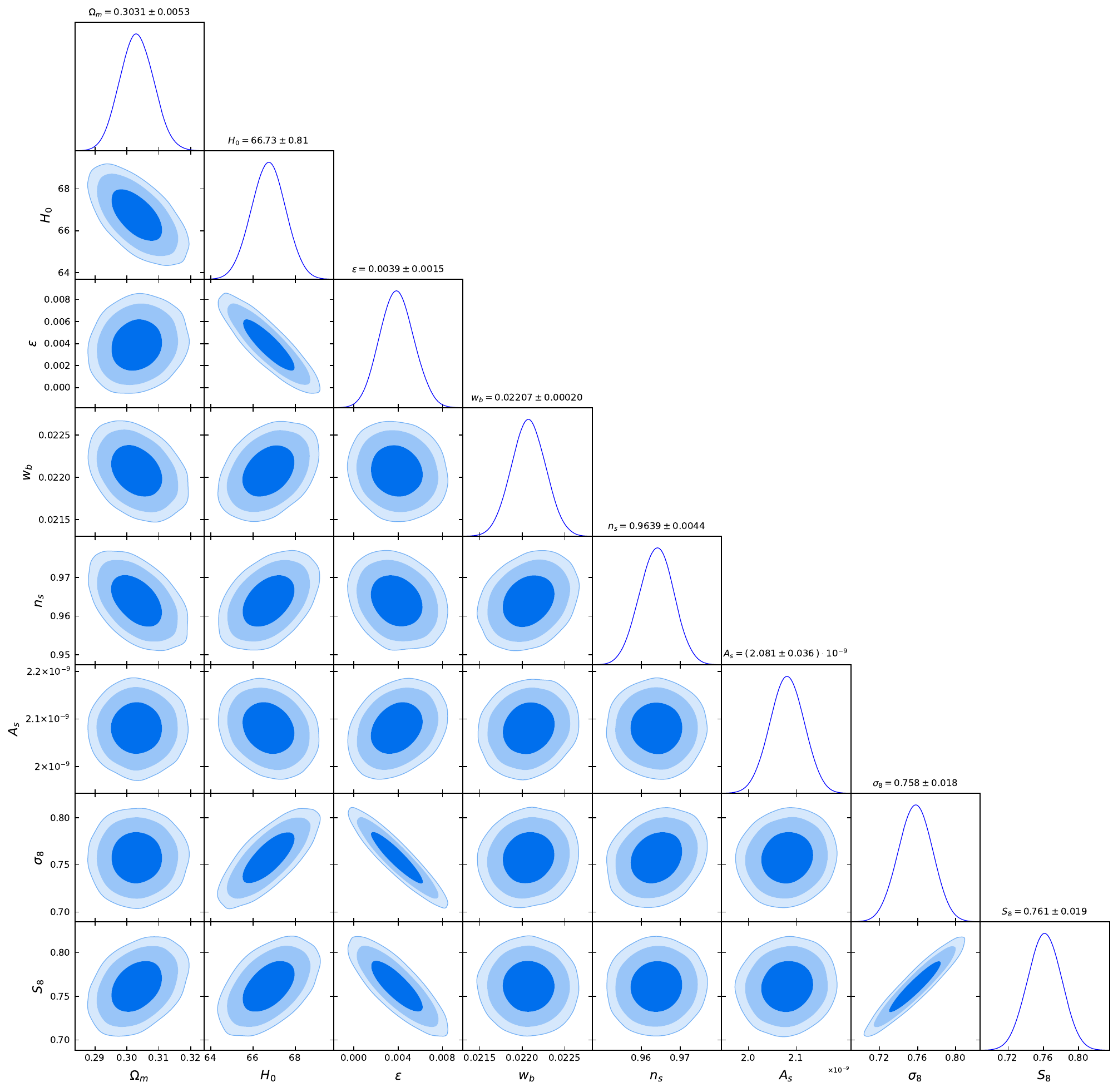}}
\caption{Same as in Fig. \ref{fig:GR_LCDM_Baseline} but for the BD-$\Lambda$CDM cosmological model.}
\label{fig:BD_LCDM_Baseline}
\end{figure*}
\begin{figure*}[htbp]
\centering
\mbox{\includegraphics[width=170mm]{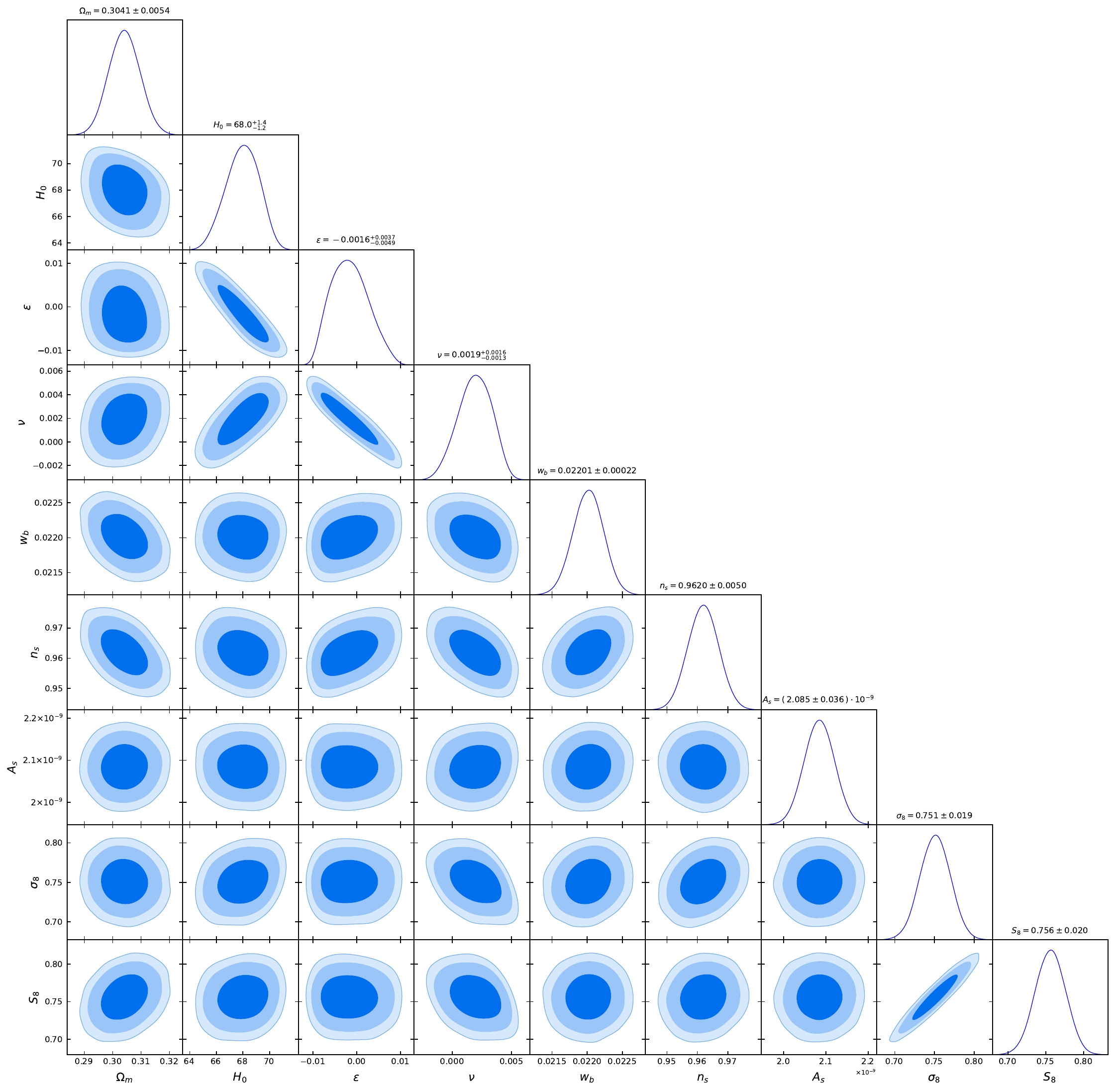}}
\caption{Same as in Fig. \ref{fig:GR_LCDM_Baseline} but for the BD-RVM cosmological model.}
\label{fig:BD_RVM_Baseline}
\end{figure*}
\begin{figure*}[htbp]
\centering
\mbox{\includegraphics[width=170mm]{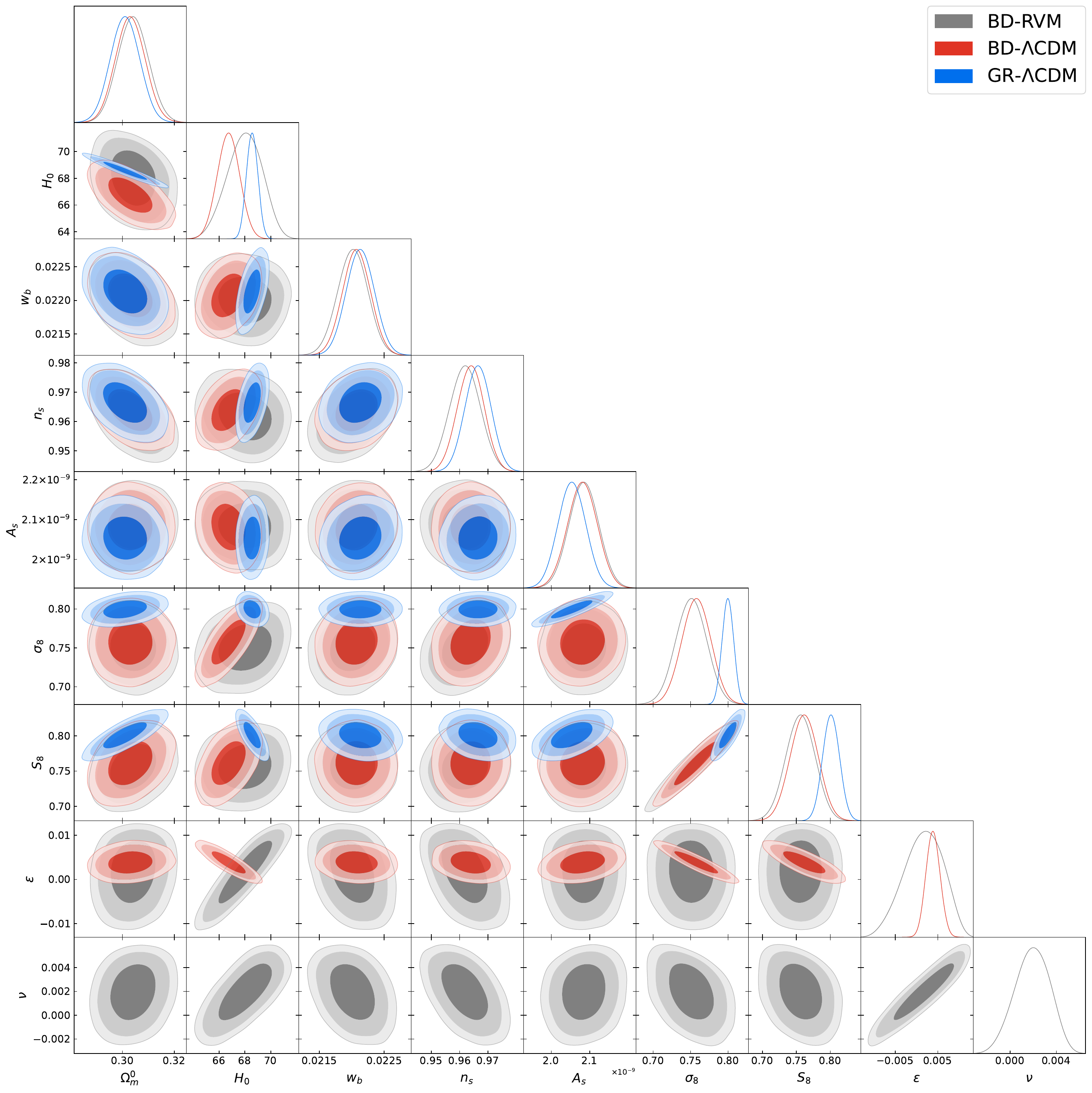}}
\caption{Two-dimensional marginalized likelihood distributions, for each one of the parameters, in the GR-$\Lambda$CDM, BD-$\Lambda$CDM and BD-RVM models at 1$\sigma$ and 2$\sigma$ c.l., obtained considering the Baseline dataset. We also included the corresponding one-dimensional marginalized likelihood distributions. The $H_0$ parameter is expressed in km/s/Mpc.}
\label{fig:Baseline_all_models}
\end{figure*}
\begin{figure*}[htbp]
\centering
\mbox{\includegraphics[width=170mm]{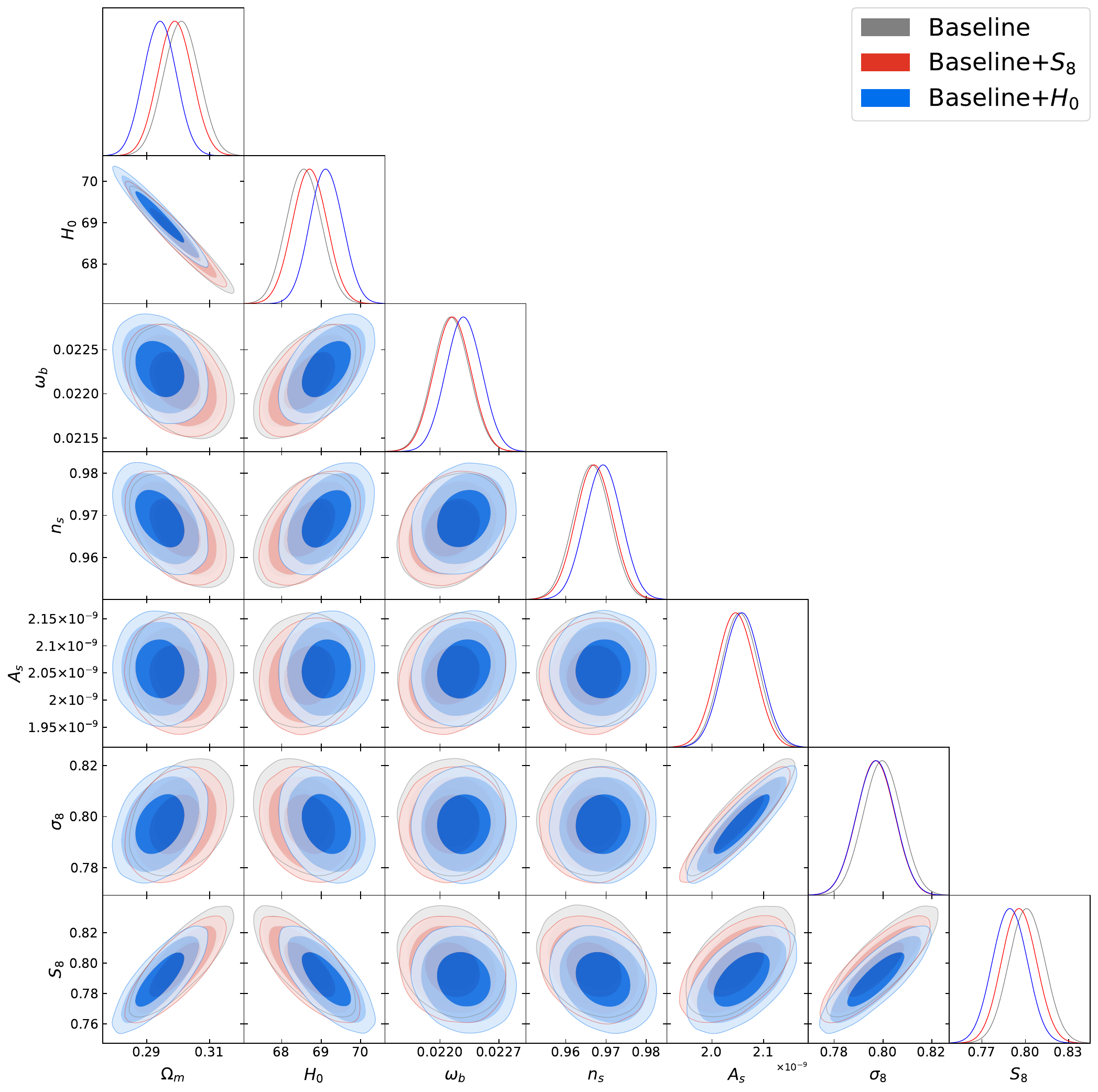}}
\caption{Two-dimensional marginalized likelihood distributions, for each one of the parameters, in the GR-$\Lambda$CDM model at 1$\sigma$, 2$\sigma$ and 3$\sigma$ c.l., obtained considering the Baseline (gray contours), the Baseline+$S_8$ (red contours) and the Baseline+$H_0$ (blue contours) datasets. We also included the corresponding one-dimensional marginalized likelihood distributions. The $H_0$ parameter is expressed in km/s/Mpc. }
\label{fig:GR_LCDM}
\end{figure*}
\begin{figure*}[htbp]
\centering
\mbox{\includegraphics[width=170mm]{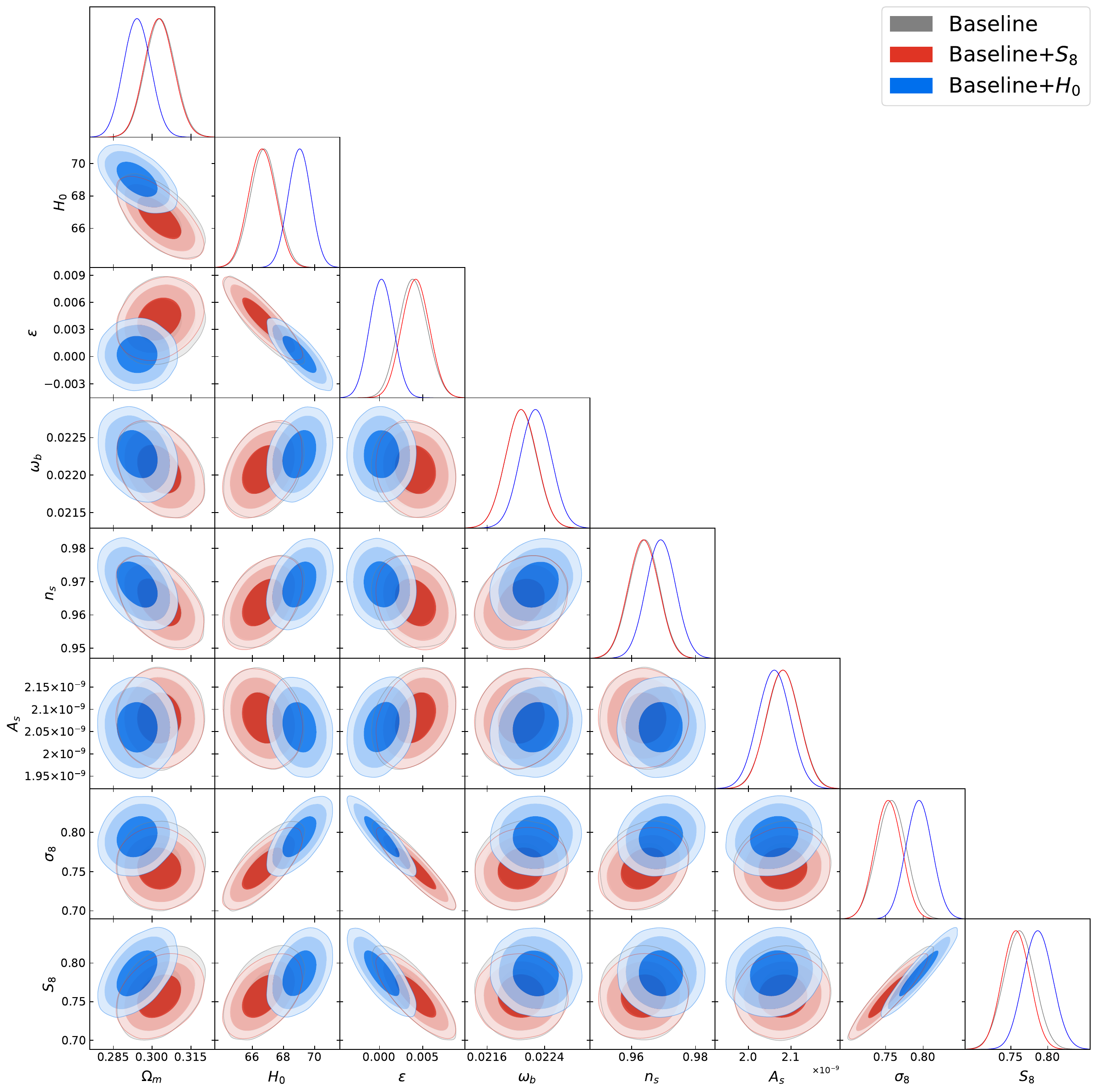}}
\caption{Same as in Fig. \ref{fig:GR_LCDM} but for the BD-$\Lambda$CDM cosmological model.}
\label{fig:BD_LCDM}
\end{figure*}
\begin{figure*}[htbp]
\centering
\mbox{\includegraphics[width=170mm]{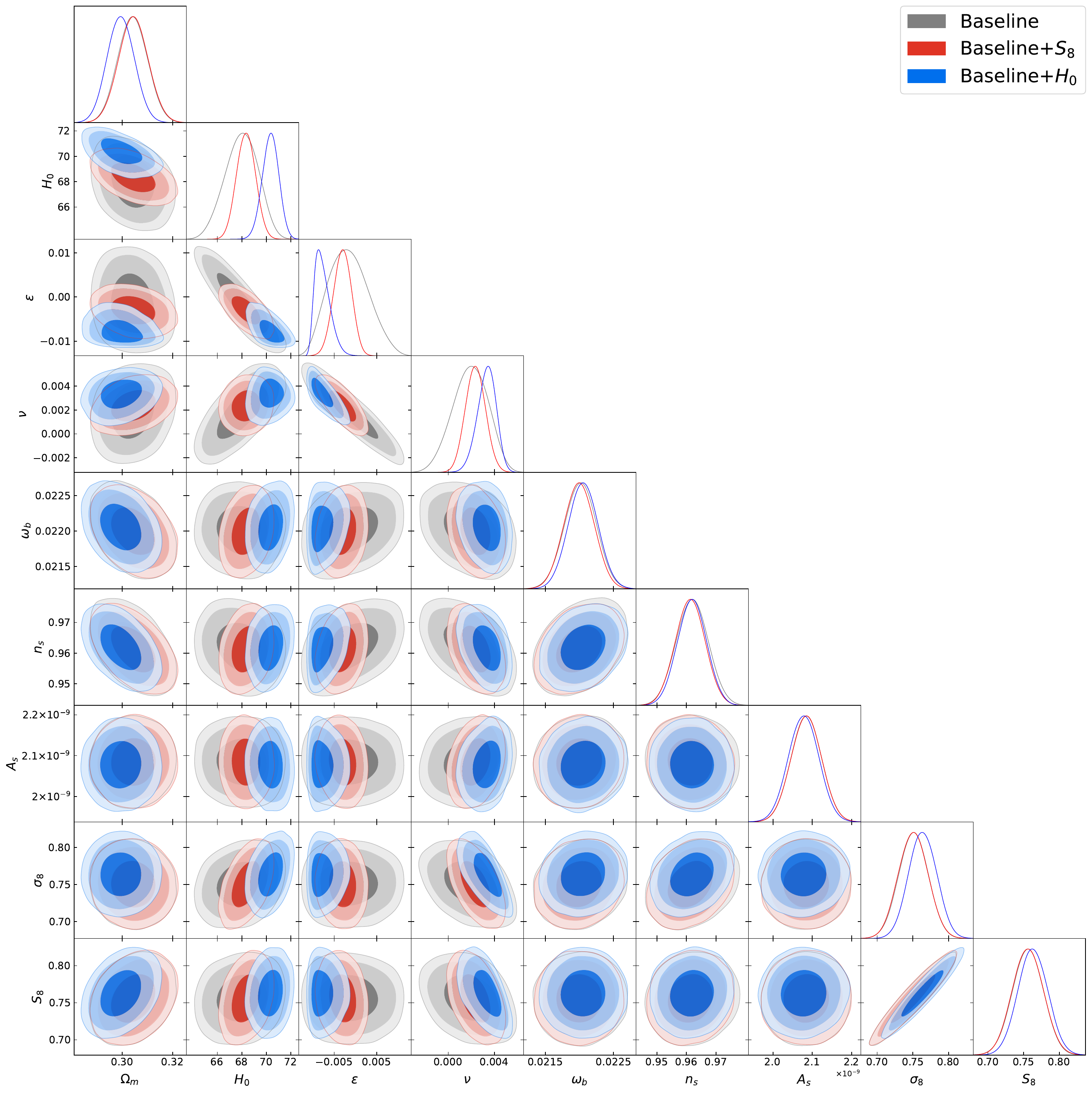}}
\caption{Same as in Fig. \ref{fig:GR_LCDM} but for the BD-RVM cosmological model.}
\label{fig:BD_RVM}
\end{figure*}
\begin{figure*}[htbp]
\centering
\mbox{\includegraphics[width=170mm]{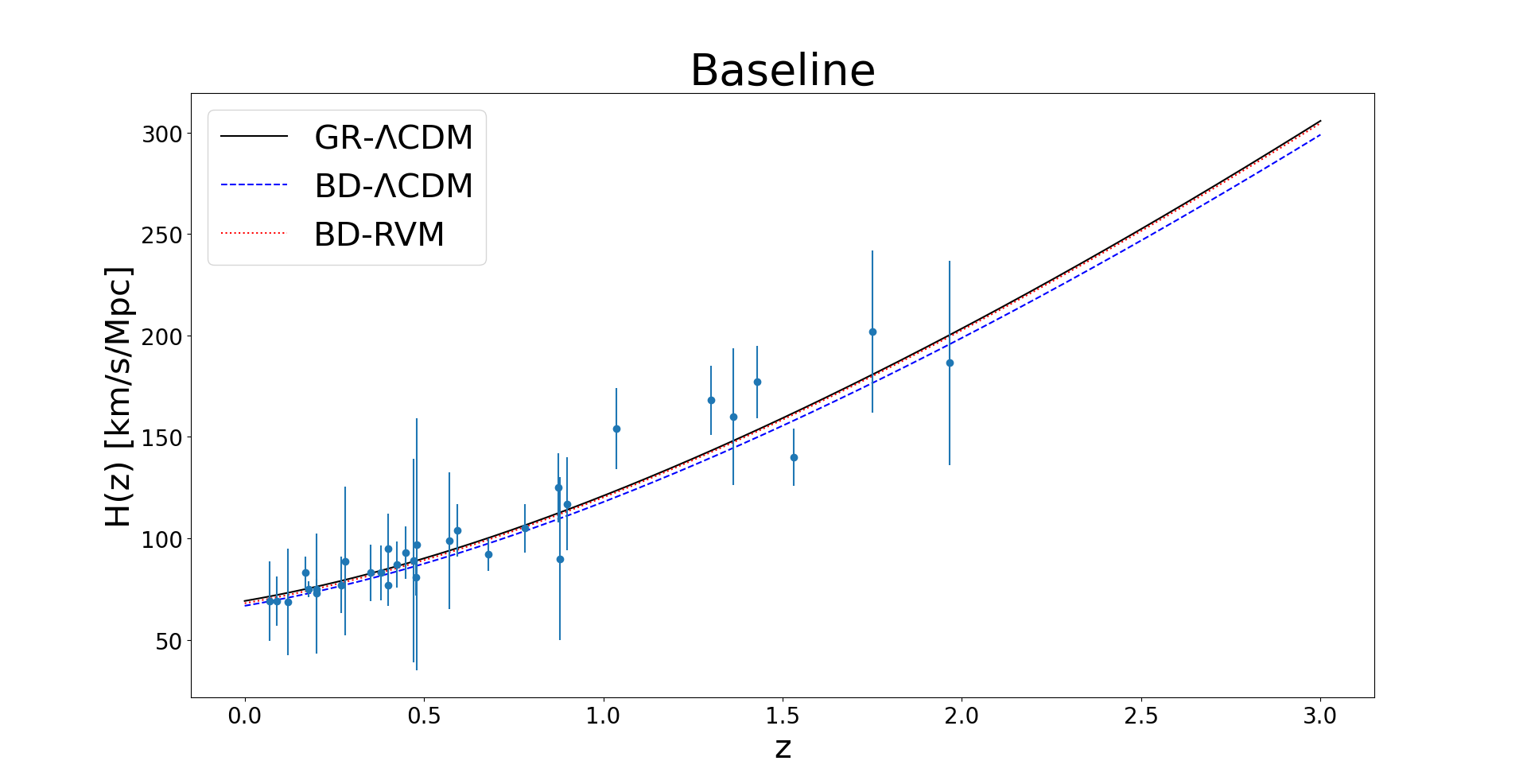}}
\caption{Evolution of the Hubble parameter as a function of the redshfit $z$ for each one of the cosmological models using the results obtained with the Baseline set.
The 33 observational data points $H(z_i)$ are shown with the corresponding error bars.  The $\chi^2_{H(z)}$ for the different models when the fitting values from Table \ref{tab:table_1} are used, are provided in Table \ref{tab:table_Baseline_contributions}. The differences among the models in fitting this particular data are not very significant as compared to other data sources. }
\label{fig:H_z_Baseline}
\end{figure*}
\begin{figure*}[htbp]
\begin{centering}
\mbox{\includegraphics[width=170mm]{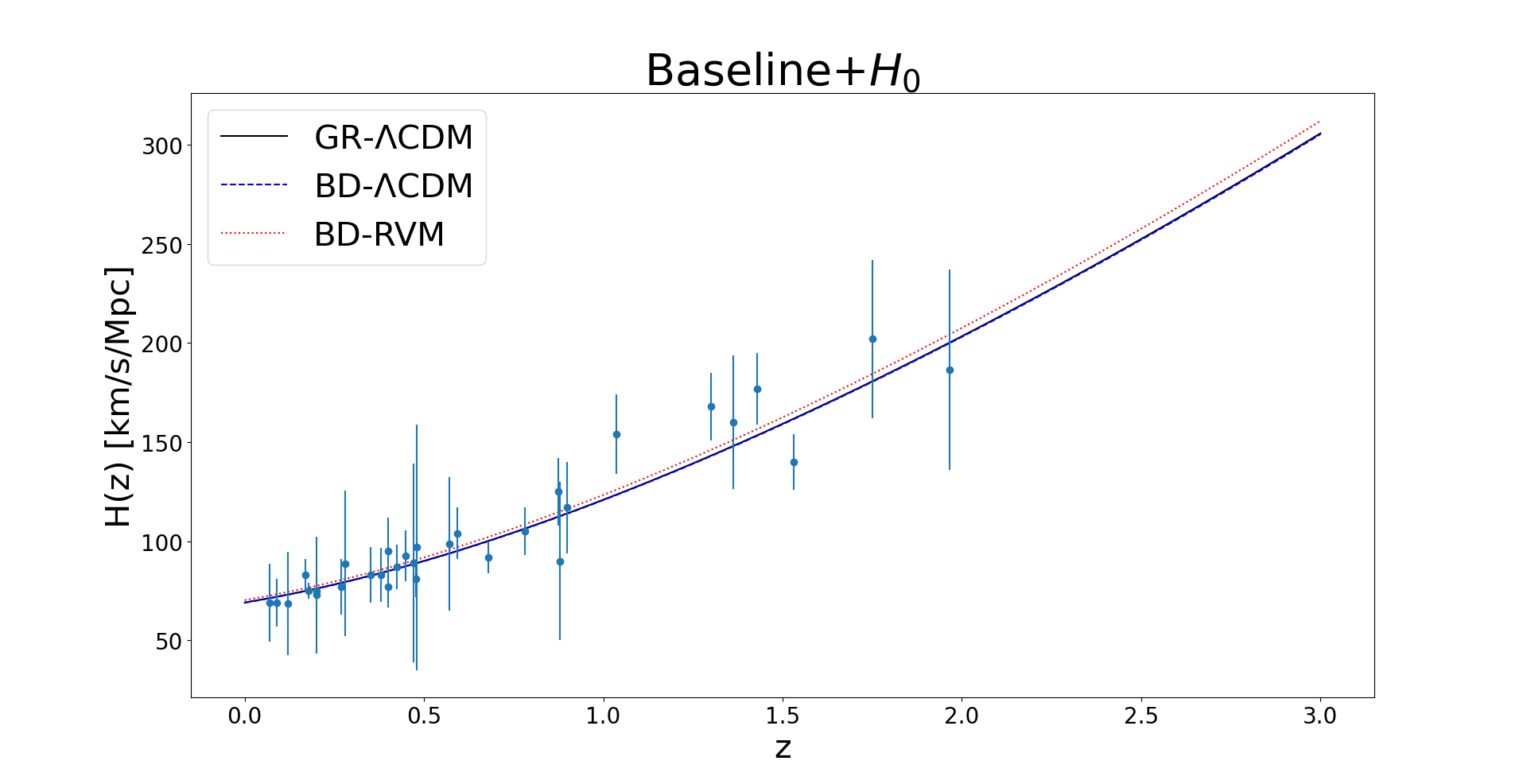}}
\end{centering}
\caption{Same as in Fig. \ref{fig:H_z_Baseline} but with the Baseline+$H_0$ set. The $\chi^2_{H(z)}$ for the different models when the fitting values from Table \ref{tab:table_3} are used, are provided in Table
\ref{tab:table_Baseline_H0_contributions}
Again, the differences among the models in fitting the $H(z)$ data are not significant.}
\label{fig:H_z_Baseline_H0}
\end{figure*}

\section{Effective EoS of the dark energy in the extended BD-framework}\label{sec:effective_EoS}
Along the lines of \cite{SolaPeracaula:2020vpg}, we now rewrite the Brans-Dicke cosmological equations in an effective GR-picture, which is useful to parameterize the departure, at the background level, of the BD theory from the standard GR-$\Lambda$CDM model. Using the field equations \eqref{eq:Friedmann_equation}-\eqref{eq:pressure_equation}-\eqref{eq:SF_equation} of  BD gravity, we find that the  Friedmann and  pressure equations of the effective GR theory describing them  take the following form:
\begin{align}
& 3H^2= 8\pi{G_N}\left(\rho_m + \rho_r + \rho_{\textrm{vac}} + \rho_\varphi\right)\\
& 2\dot{H} + 3H^2 = -8\pi{G_N}\left(p_r + p_{\textrm{vac}} + p_\varphi \right)
\end{align}
where we have introduced the dimensionless BD-field
\begin{equation}\label{eq:varphi}
\varphi\equiv G_N\phi = a^{-\epsilon}
\end{equation}
and used Eq.\,\eqref{eq:power-law_BD_field}.  Notice that in the above generalized Friedmann's and pressure equations 
we have encoded the non-GR terms of the BD theory as if they were a new fluid (the  BD-fluid) characterized by the following energy density and pressure:
\begin{align}\label{eq:rho_and_p_BD}
&\rho_\varphi \equiv \frac{3}{8\pi{G_N}}\left(H^2\Delta\varphi - H\dot{\varphi} +\frac{\wBD}{6}\frac{\dot{\varphi}^2}{\varphi} \right)\\ \label{eq:rho_and_p_BD2}
&p_\varphi \equiv \frac{1}{8\pi{G_N}}\left(-3H^2\Delta\varphi -2\dot{H}\Delta\varphi + \ddot{\varphi} + 2H\dot{\varphi} + \frac{\wBD}{2}\frac{\dot{\varphi}^2}{\varphi} \right)
\end{align}
where  we have defined 
\begin{equation}\label{eq:delta_varphi}
\Delta\varphi \equiv 1 - \varphi\,.
\end{equation}
According to \eqref{eq:varphi}, $\Delta\varphi=0$ only in the current universe.
One can check from the field equations that the effective BD-fluid \eqref{eq:rho_and_p_BD}-\eqref{eq:rho_and_p_BD2} fulfills the energy conservation equation
\begin{equation}\label{eq_BD_conservation_equation}
\dot{\rho}_\varphi + 3H\left(p_\varphi + \rho_\varphi\right) = 0.      
\end{equation}
Once we have defined the above quantities we can go a step further and compute the effective equation of state  (EoS) of the combined  BD-fluid and the vacuum energy density (VED), namely: 
\begin{align}
w_{\textrm{eff}}(z)\equiv \frac{p_{\textrm{vac}} + p_\varphi }{\rho_{\textrm{vac}} + \rho_\varphi }=-1+\frac{p_{\varphi} + \rho_\varphi }{\rho_{\textrm{vac}}+\rho_\varphi}\,,
\end{align}
where we used the fact that $p_{\textrm{vac}}=-\rho_{\textrm{vac}}$.
With the help of equations \eqref{eq:rho_and_p_BD} and \eqref{eq:rho_and_p_BD2} we find
\begin{align}
\label{eq:effective_EoS}
w_{\textrm{eff}}(z)=-1 + \frac{-2\dot{H}\Delta\varphi + f_1(\varphi,\dot{\varphi},\ddot{\varphi})}{\Lambda(H)(1-\Delta\varphi) + 3H^{2}\Delta\varphi + f_2(\varphi,\dot{\varphi})  }  
\end{align}
where
\begin{eqnarray}
&f_1(\varphi,\dot{\varphi},\ddot{\varphi}) = \ddot{\varphi} - H\dot{\varphi}  + \wBD\frac{\dot{\varphi}^2}{\varphi}\\
&f_2(\varphi,\dot{\varphi}) = -3H\dot{\varphi} + \frac{\wBD}{2}\frac{\dot{\varphi}^2}{\varphi}.\label{eq:f2}
\end{eqnarray}
We note that for the BD-$\Lambda$CDM model  we must replace $\Lambda(H)(1-\Delta\varphi)\Rightarrow \Lambda=\textrm{const.}$  in \eqref{eq:effective_EoS} since for this model, in the expression of $\rho_{\textrm{vac}}$, the term $\CC$ is constant and $\varphi=1$. The explicit form of the effective EoS in terms of the cosmological redshift, $z$, can be obtained upon combining \eqref{eq:effective_EoS}-\eqref{eq:f2} with equations \eqref{eq:Hubble_function_RVM} and \eqref{eq:varphi}.  We refrain from writing out this formula, but in order to better appreciate the behaviour of $w_{\rm eff}(z)$ close to the present time we provide the following analytic approximation in terms of  $z$, valid up to ${\cal O}(z^3)$ :
\begin{equation}
\begin{split}
w_{\textrm{eff}}(z) = -1 + \frac{\epsilon}{1-\Omega^0_m}\left[  \frac{1}{3}
+ \frac{\Omega^0_m}{2} + \frac{\wBD\epsilon}{3}\right.\nonumber
\end{split}
\end{equation}
\begin{equation}
\label{eq:effective_EoS_approximation}
\begin{split}
+\Omega^0_m\left(\frac{3}{2} +\wBD\epsilon  \right) z + \Omega^0_m{\wBD\epsilon}z^2 \nonumber
\end{split}
\end{equation}
\begin{equation}
\label{eq:effective_EoS_approximation}
\begin{split}
 \left. +\Omega^0_m\left(1 + \frac{\wBD\epsilon}{3}\right)z^3    \right] + \mathcal{O}\left(\epsilon^2,\epsilon\nu,z^4\right)\,.
\end{split}
\end{equation}
where we neglect the higher order powers of the parameters $\epsilon$ and $\nu$ beyond the linear order since they both satisfy $|\epsilon,\nu|\ll1$. However, we keep  the product $\wBD\epsilon$ since it can be of order $1$ for large enough $\wBD$.  Similarly $\wBD\epsilon^2\sim \epsilon$ is also retained as any term of ${\cal O} (\epsilon)$.

The effective EoS of the BD-fluid +VED evolves with the expansion, and when viewed from the GR standpoint it describes an effective dark energy (DE) inherent to the BD theory with cosmological term. This was studied in \cite{SolaPeracaula:2019zsl,SolaPeracaula:2020vpg} for the case of $\CC=$const., which is the situation with the model BD-$\CC$CDM being studied here,  but in this work we extend the analysis also for the case BD-RVM, which is  when $\CC$ is actually `running'  within the context of the RVM, see Eq.\,\eqref{eq:LambdaRVM}.
The plots of the effective EoS \eqref{eq:effective_EoS} as a function  of the redshift for the extended BD frameworks under study are             depicted in Figs. \ref{fig:w_eff_Baseline} and \ref{fig:example}.
%

If we consider the above expression for today's universe ($z=0$) and  utilize the relation \eqref{eq:relation_wBDeps}, without including the (currently very small) radiation contribution, we get 
\begin{equation}w_{\textrm{eff}}(0) = -1 - \epsilon{F(\Omega^0_m)} + \mathcal{O}\left(\epsilon^2,\epsilon\nu\right)
\end{equation} with
\begin{equation}
F(\Omega^0_m) = \frac{4-10\,\Omega^0_m +3 \Omega^{0^{2}}_m }{12 -18 \Omega^0_m + 6\Omega^{0^{2}}_m }\,.
\end{equation}
We point out the following interesting feature of the previous result:  being $F(\Omega^0_m)>0$ for the current values of the cosmological parameters, the behavior of the EoS at $z=0$ turns out to be phantom-like or quintessence-like depending on the sign of $\epsilon$. The BD-RVM mimics quintessence (since $\epsilon<0$ for it in all fitting scenarios considered here, see  Tables \ref{tab:table_1}, \ref{tab:table_2} and \ref{tab:table_3}), whereas BD-LCDM mimics phantom-like behavior (since $\epsilon>0$ for this model in the same fitting tables).  It is worth noticing that quintessence and phantom like behaviors can effectively be  mimicked by extensions of BD gravity which do not involve any sort of  ad hoc scalar fields but just rigid or dynamical vacuum effects.

\section{Structure Formation}\label{sec:structure_formation}
In order to properly test the theoretical predictions of the various cosmological models under study we must consider structure formation. The differential equation for the matter density contrast $\delta_m\equiv \delta\rho_m/\rho_m$ for the models considered here can be approximated as follows\,\cite{deCruzPerez:2018cjx}:
\begin{equation}\label{eq:structure_formation_equation}
\delta^{\prime\prime}_m +\left(3 + a\frac{H^{\prime}(a)}{H(a)}\right)\frac{\delta^{\prime}_m}{a} -\frac{4\pi{G(a)}}{H^2(a)}\frac{\delta_m}{a^2} = 0\,,     
\end{equation}
where $()^{\prime}\equiv d/da$ denotes derivative with respect to the scale factor. The Hubble function is the corresponding one for each model. Notice also that  $G(a)=1/\phi(a)=G_N a^\epsilon$ is given by Eq.\,\eqref{eq:power-law_BD_field}, and in particular $G(a)=G_N$ for the GR-$\Lambda$CDM model ($\epsilon=0$).  The above approximation for the structure formation equation turns out to be accurate since the main effects actually come from the different expressions of the Hubble function in each model. 

We are interested in finding the required initial conditions deep into the matter dominated epoch, which means that we calculate the limit of the different terms in \eqref{eq:structure_formation_equation} when the scale factor $a$ is small enough, see \cite{deCruzPerez:2018cjx} for the details. In that limit the resultant differential equation admits power law solutions (we only consider the growing mode solution) $\delta_m\sim a^s$. For each model we find: 
\begin{align}\label{eq:initial_conditions}
&{\bf GR-\Lambda{CDM}}:\quad s= 1 \\
&{\bf BD-\Lambda{CDM}}:\quad s\simeq 1 - \frac{4}{5}\epsilon -\frac{1}{10}\epsilon^2\wBD \\
&{\bf BD-RVM}:\quad s \simeq 1 - \frac{4}{5}\epsilon - \frac{6}{5}\nu -\frac{1}{10}\epsilon^2\wBD , 
\end{align}
where for the BD-$\Lambda$CDM we have neglected the terms of order $\mathcal{O}(\epsilon^2)$ whereas for the BD-RVM we have neglected those of order $\mathcal{O}(\epsilon^2,\epsilon\nu,\nu^2)$.
Notice that in order to have suppression of structure formation we need that the overall contribution from the $\epsilon$, $\nu$ and $\epsilon^2\wBD$ provides $s<1$. \textcolor{black}{The suppression of the structure formation with respect to the GR-$\Lambda$CDM model not only comes from the initial conditions but also from the modification of the terms in Eq.\,\eqref{eq:structure_formation_equation}. As we shall see, this is why both models, BD-$\Lambda$CDM and BD-RVM, can improve the fitting of the LSS data when compared with the standard model.  \jtext{ In Ref.\cite{SolaPeracaula:2020vpg} (cf. appendices C and D) one can see that Eq.\,\eqref{eq:structure_formation_equation} emerges as a very good approximation for the study of the matter perturbations in the BD framework within the horizon.  This is sufficient for the present analysis. The reader may consult that study in order to see how exactly structure formation becomes suppressed in the context of the BD-$\Lambda$CDM model. Additionally, in the works \cite{Gomez-Valent:2018nib,Gomez-Valent:2017idt} it is assessed in great detail the impact of running vacuum on $f\sigma_8$ data. The reason why the BD-RVM model can ease the $S_8$-tension is based on the same  underlying mechanism described in these references. }}

Comparing with the available structure formation data requires the computation of the standard LSS quantity $f(z)\sigma_8(z)$, where $f(z)=d\ln\delta_m(a)/d\ln{a}$ is usually called the growth factor and $\sigma_8(z)$ is the rms mass fluctuations on $R_8=8{h^{-1}}$ Mpc scales, whose expression is: 
\begin{equation}\label{eq:sigma8_expression}
\sigma^2_8(z) = \delta^2_m(z)\int \frac{d^{3}k}{(2\pi)^3}P(k,\vec{p})W^2(k{R_8}). 
\end{equation}
$P(k,\vec{p}) = P_{0}k^{n_s}T^2(k,\vec{p})$ is the linear matter power spectrum, $P_0$  a normalization constant (see below), $T(k,\vec{p})$ represents the transfer function and, finally, $\vec{p}$ a parameter vector involving the characteristic free parameters of each model. Among the options available for the transfer function, we use here the BBKS one\,\cite{Bardeen:1985tr}: 
\begin{equation}\label{eq:Transfer_function}
\begin{array}{ll}
T(x) = &\frac{\ln (1+0.171 x)}{0.171\,x}\Big[1+0.284 x + (1.18 x)^2+\\
   & + \, (0.399 x)^3+(0.490x)^4\Big]^{-1/4}\,.
\end{array}
\end{equation}
In the above expression $x\equiv k/\tilde{\Gamma}k_{\textrm{eq}}$ where 
\begin{equation}
k_{\textrm{eq}} = a_{\textrm{eq}}H(a_{\textrm{eq}})    
\end{equation}
is the value of the comoving wave number at the equality point between matter and radiation, defined through $\rho_r(a_{\textrm{eq}})=\rho_m(a_{\textrm{eq}})$. The quantity $\tilde{\Gamma}=e^{-\Omega^0_b-\sqrt{2h}\frac{\Omega^0_b}{\Omega^0_m}}$ is known as the modified shape parameter and it incorporates the effect of baryons in the transfer function. In \eqref{eq:sigma8_expression} we have also introduced the top-hat smoothing function 
\begin{equation}\label{eq:smoothing_function}
W(k{R_8}) = \frac{3}{k^{2}R^2_8}\left(\frac{\textrm{sin}(k{R_8})}{k{R_8}} -\textrm{cos}(k{R_8})\right)\,,   
\end{equation}
with $R_8$ defined previously. There are several ways of dealing with the normalization factor $P_0$ of the matter power spectrum. In this work, the procedure followed takes advantage of using the connection between $P_0$ and the normalization factor of the primordial power spectrum, $A_s$, for which we have obsevational constraints. The relation between these quantities is  \cite{Sola:2017jbl}:  
\begin{equation}\label{eq:P0}
P_0=A_s\frac{8\pi^2}{25}\frac{k_*^{1-n_s}}{(\Omega^0_m h^2)^2(100\varsigma)^4}\,,
\end{equation}
where $k_*=0.05\textrm{Mpc}^{-1}$ is the pivotal scale used by the Planck team and $\varsigma\equiv 1$ km/s/Mpc$=2.1332\times10^{-44} \textrm{GeV}$. In contrast to \cite{Sola:2017jbl}, however, here we can use $A_s$ as a fit parameter since we computed a compressed likelihood involving this parameter with the remaining distance priors, see \eqref{eq:correlationCMB} below.  To ease the computation what we actually do is calculate the normalized quantity $\sigma_8(z)/\sigma^{\Lambda}_8(0)$, where $\sigma^{\Lambda}_8(0)=0.8118$ is the value obtained by the Planck team for the analysis Planck 2018 TT+lowE \cite{Planck:2018vyg} with the vanilla model. Therefore, taking into account the definitions of the different quantities involved we may compute the value of the $\sigma_8(z)$ function at different redshifts as follows:
\begin{equation}\label{eq:normalized_sigma8}
\sigma_8(z) = \sigma^{\Lambda}_8(0)\sqrt{\frac{P_0}{P^{\Lambda}_0}}\frac{\delta_m(z)}{\delta^{\Lambda}_m(0)}\frac{\gamma}{\gamma_\Lambda}
\end{equation}
where for the sake of simplicity we have defined 
\begin{equation}\label{eq:gamma}
\gamma \equiv \left(\int^{\infty}_{0}dk k^{2 + n_s}T^2(k,\vec{p})W^2(kR_8) \right)^{1/2}\,.
\end{equation}
In order to fix the normalization constant $P_0$ we have to choose a fiducial cosmology. In this work we choose the vanilla standard model of cosmology constrained with Planck 2018 TT+lowE data. So, in \eqref{eq:normalized_sigma8} those quantities labeled with $\Lambda$ ( superscript or subscript) correspond to either the fiducial values, of the cosmological parameters, found in the Planck 2018 TT+lowE analysis \cite{Planck:2018vyg} or to the values of the functions (\eqref{eq:Transfer_function}-\eqref{eq:P0} and \eqref{eq:gamma}) computed using these fiducial parameters. Similarly, the quantity denoted by $\delta^{\Lambda}_m(0)$ corresponds to the present value of the matter density contrast for the fiducial model obtained after solving the perturbations equation \eqref{eq:structure_formation_equation}.  
\newline
Thus, using \eqref{eq:normalized_sigma8} for $\sigma_8(z)$  and the expression for the growth factor $f(z)=d\ln\delta_m(a)/d\ln{a}$ which follows upon solving \eqref{eq:structure_formation_equation} for the density contrast $\delta_m(a)$ for each model, we are in position to numerically compute the values of the observable $f(z)\sigma_8(z)$ at different redshifts.

\section{Data and methodology}\label{sec:Data_and_methodology}
We fit the GR-$\Lambda$CDM, the BD-$\Lambda$CDM and the BD-RVM cosmological models to a very complete and robust dataset compiled from: distant type Ia Supernovae (SNIa), baryonic acoustic oscillations (BAO), a compilation of measurements of the Hubble parameter at different redshifts ($H(z_i)$), large-scale structure formation data ($f(z_i)\sigma_8(z_i)$), CMB Planck 2018 TT+lowE data, a prior on the weak-lensing observable ($S_8$) and, finally, a prior on the $H_0$ parameter \footnote{The dataset employed in this work is very similar to the one used in references \cite{SolaPeracaula:2020vpg} and \cite{SolaPeracaula:2021gxi} check them for more details.}. In what follows a brief description of each one of the datasets is provided, together with the corresponding references:
\newline
\newline
{\bf SNIa}: 6 effective points on the inverse of the normalized Hubble function $1/E(z_i)$, with its corresponding covariance matrix, from the Pantheon+MCT sample \cite{Pan-STARRS1:2017jku,Riess:2017lxs}, which compresses the information from 1063 SNIa.
\newline
\newline
{\bf BAO}: We employ data on both, isotropic and anisotropic BAO estimators \cite{Carter:2018vce,Kazin:2014qga,Gil-Marin:2016wya,duMasdesBourboux:2020pck,DES:2017rfo,Neveux:2020voa}. See \cite{SolaPeracaula:2020vpg} and \cite{SolaPeracaula:2021gxi} in order to see the tables with the data points.  
\newline
\newline
\textbf{H(z)}: We use the 32 data points on the Hubble parameter $H(z_i)$ \cite{Jimenez:2003iv,Simon:2004tf,Stern:2009ep,Moresco:2012jh,Zhang:2012mp,Moresco:2015cya,Moresco:2016mzx,Ratsimbazafy:2017vga,Borghi:2021rft,Favale:2023lnp}, all of them, obtained employing the differential age technique, and therefore defining a robust type of $H(z)$ data which are not correlated with BAO,  the co-called cosmic chronometers.
\newline
\newline
{\bf LSS}: 14 large scale structure (LSS) data points at different redshifts embodied in the observable $f(z_i)\sigma_8(z_i)$ \cite{Said:2020epb,Blake:2013nif,Simpson:2015yfa,Gil-Marin:2016wya,Blake:2011rj,Mohammad:2018mdy,Guzzo:2008ac,Song:2008qt,Neveux:2020voa}.
\newline
\newline
{\bf CMB}: From Planck 2018 TT+lowE data \cite{Planck:2018vyg} we have computed a compressed likelihood for the distance priors $\mathcal{R}$ (shift parameter) and $\ell_a$ (acoustic length) together with the corresponding correlations for the parameters ($\omega_b,n_s,A_s$). \textcolor{black}{With the aim of comparing with previous works \cite{deCruzPerez:2018cjx,SolaPeracaula:2019zsl,SolaPeracaula:2020vpg} we do not include the Planck 2018 polarization data.} \footnote{\joan{This is not an undue limitation of our analysis. The Planck 2018 TT+lowE data are perfectly sound to test new models; and, as we indicated previously, for the sake of a fairer comparison  of the current analysis with the aforementioned works it is perfectly sound to use such a valuable CMB data. In addition, let us note that the polarization data is still under analysis and is not completely free from moderate controversies in the literature, see e.g. \cite{Garcia-Quintero:2019cgt,Lin:2017bhs}. While we do not judge at all these discussions we would like that our models are tested against the cleanest and nevertheless equally robust CMB data with which we already had tested similar models before.}}.  The mean values and the corresponding error bars for the different components are $\omega_b = 0.022038\pm 0.00022$, $A_s = (2.0898\pm 0.034)\times 10^{-9}$, $n_s = 0.9641\pm 0.0058$, $\ell_a=301.54\pm 0.1319$ and  $\mathcal{R}=1.7469\pm 0.0072$, whereas the matrix containing the correlations between the different elements takes the form:
\begin{small}
\begin{equation}\label{eq:correlationCMB}
C_{\textrm{CMB}} = 
\begin{pmatrix}
1 & 0.1193 & 0.4926 & -0.3014 & -0.5546  \\
0.1193 & 1 & 0.0262 & -0.0434 & 0.0240  \\
0.4926 & 0.0262 & 1 & -0.3571 & -0.7776   \\
-0.3014 & -0.0434 & -0.3571 & 1 & 0.4207  \\
-0.5546 & 0.0240 & -0.7776 & 0.4207 & 1 
\end{pmatrix}
\end{equation}
\end{small} 
\newline
\newline
{\bf Prior on $S_8$}: A prior value on the observable $S_8\equiv\sigma_8\sqrt{\Omega^0_m/0.3} = 0.737^{+0.040}_{-0.036}$ from the gravitational weak-lensing survey KiDS+VIKING-450 \cite{Hildebrandt:2018yau}. 
\newline
\newline
{\bf Prior on $H_0$}: We consider the prior  on the $H_0$ parameter provided by the SH0ES team \cite{Riess:2021jrx} (obtained from the cosmic distance ladder method), namely $H_0=73.04\pm 1.04$ km/s/Mpc. This measurement exceeds the Planck$+\Lambda$CDM result by $5\sigma$ (one in 3.5 million), making it implausible to reconcile the two by chance. \textcolor{black}{By including this prior in the analysis we just want to test effectively whether the models under study are able to increase the value of the $H_0$ parameter without spoiling the fit to the other data sets. This strategy was actually applied in former works of us\,\cite{deCruzPerez:2018cjx,SolaPeracaula:2019zsl,SolaPeracaula:2020vpg,Singh:2012zzd,SolaPeracaula:2021gxi} and here we want to verify if the inclusion of the running vacuum helps to ease the $H_0$-tension. It is important to note that it is the consideration of the $H_0$ prior what forces (if that is possible at all for a given model) the boost in the value of that parameter. We can see that for the Baseline data set the overlap between the  fitted value  of $H_0$ and the value of the SH0ES measurement does not come naturally.  However, for the Baseline+$H_0$ data set the BD-RVM model shows immediately a clear predisposition to enhance the value of $H_0$ in contrast to the other models. } \joan{We  believe that our procedure, which as indicated was used in previous works, can be useful to effectively test the ability of different models to accommodate higher values of $H_0$. Given a model, the success is in general not granted and in fact it does not even work  with generic parameterizations of the DE, such as the well-known XCDM parameterization (also called $w$CDM), as can be checked e.g. in the comparative numerical study presented in \cite{SolaPeracaula:2020vpg}. When it does, however,  it then becomes suggestive that the model may be a promising framework  for relieving the tension.}
\newline
\newline
We do not use simultaneously all of the data points given above to fit the models under study, but arrange for three  different combinations of observational points. Specifically, our dataset scenarios are the following: 
\begin{itemize}
\item {\bf Baseline}: We consider as our Baseline dataset the one made of the following string of data: SNIa+$H(z)$+BAO+LSS+CMB. For this scenario, the SH0ES prior for $H_0$ is not included. 
\item {\bf Baseline+$S_8$}: In addition to the data considered in the Baseline set, we add the prior on the $S_8$ parameter. 
\item {\bf Baseline+$H_0$}: The Baseline data is in this case complemented with the prior on the $H_0$ parameter provided by the SH0ES team. 
\end{itemize}
\jnew{Let us make clear at this point that the additional enhancement that might be obtained in the cases where a prior is added on top of the Baseline data (such as the situation in the last two dataset scenarios) is to be interpreted with care since the inclusion of priors may artificially drag the models' performance towards the desired direction. This is not always so, but it may happen.  See  Sections \ref{sec:discussion} and \ref{sec:conclusions} for further discussions on the impact of priors in the present analysis. Our main conclusions, however, will be derived from the results obtained with the Baseline dataset free from priors. The results obtained from the datasets containing a prior will be taken only as indicative.}

Using the above datasets,  we proceed to analyze whether  the two extensions of the  Brans \& Dicke cosmology being considered in this work, namely the BD-$\Lambda$CDM and  BD-RVM models,  are capable to fit the observations in a comparable way to the concordance  model (GR-$\CC$CDM), or maybe even better, and in particular if they are able to cope  with the $\sigma_8$-tension \cite{DiValentino:2020vvd}  and the $H_0$-tension \cite{DiValentino:2020zio}. 
\begin{table*}
\begin{ruledtabular}
\begin{tabular}{cccc}
\\[+0.5mm]
\multicolumn{1}{c}{} & \multicolumn{3}{c}{\Large Baseline}
\\[+0.5mm]
\\\hline
\\[+1mm]
 & { \bf GR-$\Lambda$CDM}  & { \bf BD-$\Lambda\textrm{CDM}$ }   & { \bf BD-$\textrm{RVM}$}\\  
\\[+1mm]
{ $\chi^2_{\textrm{SNIa}}$}  & { $5.57$}  & { $5.62$}    & { $5.63$}\\
{$\chi^2_{\textrm{BAO}}$}   & {$13.33$} & {$14.26$} & {$14.49$}\\
{$\chi^2_{H(z)}$}    & {$14.69$} & {$16.47$}  & { $14.92$}\\
{$\chi^2_{f\sigma_8}$}   & {$16.46$} & { $9.78$}  & { $9.27$}\\  
{$\chi^2_{\textrm{CMB}}$}   & { $1.78$} & { $0.19$}   & { $0.43$}\\  
\end{tabular}
\\[+1mm]
\caption{\label{tab:table_Baseline_contributions} We display the contributions of some of the data employed to test the cosmological models, obtained considering the mean values provided in Table \ref{tab:table_1} for the Baseline data set. } 
\end{ruledtabular}
\end{table*}

\begin{table*}
\begin{ruledtabular}
\begin{tabular}{cccc}
\\[+0.5mm]
\multicolumn{1}{c}{} & \multicolumn{3}{c}{\Large Baseline+$S_8$}
\\[+0.5mm]
\\\hline
\\[+1mm]
 & { \bf GR-$\Lambda$CDM}  & { \bf BD-$\Lambda\textrm{CDM}$ }   & { \bf BD-$\textrm{RVM}$}\\  
\\[+1mm]
{ $\chi^2_{\textrm{SNIa}}$}  & { $5.54$}  & { $5.60$}    & { $5.63$}\\
{$\chi^2_{\textrm{BAO}}$}   & {$13.61$} & {$14.47$} & {$14.31$}\\
{$\chi^2_{H(z)}$}    & {$14.70$} & {$16.66$}  & { $14.73$}\\
{$\chi^2_{f\sigma_8}$}   & {$15.45$} & { $9.30$}  & { $9.15$}\\  
{$\chi^2_{\textrm{CMB}}$}   & { $2.60$} & { $0.38$}   & { $0.68$}\\
{$\chi^2_{S_8}$}   & { $2.40$} & { $0.28$}   & { $0.01$}\\ 
\end{tabular}
\\[+1mm]
\caption{\label{tab:table_Baseline_S8_contributions} \joan{Same as in Table \ref{tab:table_Baseline_contributions} but for the Baseline+$S_8$ data set. } }
\end{ruledtabular}
\end{table*}

\begin{table*}
\begin{ruledtabular}
\begin{tabular}{cccc}
\\[+0.5mm]
\multicolumn{1}{c}{} & \multicolumn{3}{c}{\Large Baseline+$H_0$}
\\[+0.5mm]
\\\hline
\\[+1mm]
& { \bf GR-$\Lambda$CDM}  & { \bf BD-$\Lambda\textrm{CDM}$ }   & { \bf BD-$\textrm{RVM}$}\\  
\\[+1mm]
{ $\chi^2_{\textrm{SNIa}}$}  & { $5.53$}  & { $5.53$}    & { $5.53$}\\
{$\chi^2_{\textrm{BAO}}$}   & {$14.32$} & {$14.29$} & {$14.90$}\\
{$\chi^2_{H(z)}$}    & {$14.73$} & {$14.75$}  & { $15.23$}\\
{$\chi^2_{f\sigma_8}$}   & {$14.45$} & { $14.24$}  & { $9.27$}\\  
{$\chi^2_{\textrm{CMB}}$}   & { $2.87$} & { $2.94$}   & { $0.51$}\\ 
{$\chi^2_{H_0}$}   & { $14.13$} & { $14.94$}   & { $6.74$}\\  
\end{tabular}
\\[+1mm]
\caption{\label{tab:table_Baseline_H0_contributions} Same as in Table \ref{tab:table_Baseline_contributions} but for the Baseline+$H_0$ data set. Notice in particular the low values of $\chi^2_{f\sigma_8}$, $\chi^2_{\textrm{CMB}}$ and $\chi^2_{H_0}$ in the case of the BD-RVM, which helps to understand why this model can alleviate the two tensions $\sigma_8$ and $H_0$ at a time. In contrast, the BD-$\CC$CDM model fails to reduce these tensions, despite it helps to lessen the $\sigma_8$ tension when the prior on $H_0$ is not included, see Table \ref{tab:table_Baseline_contributions} .} 
\end{ruledtabular}
\end{table*}

In order to compare the BD-$\Lambda$CDM and the BD-RVM with the GR-$\Lambda$CDM, we define the joint $\chi^2$-function, with Gaussian errors, as follows:
\begin{equation}\label{eq:chi2_total}
\chi^2_{\textrm{tot}}= \chi^2_{\textrm{SNIa}} +\chi^2_{H}+\chi^2_{\textrm{BAO}}+ \chi^2_{\textrm{LSS}}+\chi^2_{\textrm{CMB}}. 
\end{equation}
The different contributions of $\chi^2_{\textrm{tot}}$ are defined in the standard way. Only a few observations are in order.  In some cases such as e.g. the CMB, see Eq.\,\eqref{eq:correlationCMB}, the correlation matrix and hence the covariance matrix for $\chi^2_{\textrm{CMB}}$, is non-diagonal\,\footnote{The correlation matrix for the case with CMB polarizations and lensing is different from Eq.\,\eqref{eq:correlationCMB} and  is provided in the appendix, as well as a summary of the corresponding numerical results.}. Notice also that the term $\chi^2_H$ may contain or not the $H_0$-prior from \cite{Riess:2021jrx} depending on the dataset used for our analysis, and similarly the term $\chi^2_{\textrm{LSS}}$ may contain or not the $S_8$-prior from \cite{Hildebrandt:2018yau}.

We employ the aforementioned datasets to constrain the cosmological parameters of the models under consideration. We are also interested in determining which of the models is the one that performs better when it comes to reproducing the observational data. To do so, we utilize two statistical criteria, the Akaike information criteria (AIC) \cite{Akaike} and the deviance information criteria (DIC) \cite{DIC}, which take into account the presence of extra {\it d.o.f} by appropriately increasing the value of $\chi^2_{\textrm{min}}$. See \cite{Liddle:2007fy}, for instance, for a summarized discussion on how to use and interpret these criteria in the cosmological context.

The formula for the AIC turns out to be
\begin{equation}\label{eq:AIC}
\textrm{AIC} = \chi^2_{\textrm{min}} + \frac{2nN}{N-n-1}\,,  
\end{equation}
where $\chi^2_{\textrm{min}}$ is the minimum value of the $\chi^2$-function, $n$ represents the number of independent parameters and $N$ is the total number of data points. The second term on the \textit{r.h.s.} obviously represents the penalty for the additional number of free parameters ({\it d.o.f}).

On the other hand, the DIC value can be computed from 
\begin{equation}\label{eq:DIC}
\textrm{DIC} = \chi^2(\bar{\theta}) + 2p_D\,,    
\end{equation}
with  $p_D = \overline{\chi^2} - \chi^2(\bar{\theta})$ the effective number of parameters (compare with \eqref{eq:AIC} in the limit of large $N$),  $2p_D$ being the so-called `model complexity', which is the quantity used in the DIC criterion to  penalize the presence of extra {\it d.o.f}. The term $\overline{\chi^2}$ is the mean value of the $\chi^2$-function obtained from the Markov chains and $\bar{\theta}$ represents the mean values of the fitting parameters. 

We choose to define the AIC and DIC differences with respect to the vanilla model (GR-$\Lambda$CDM):
\begin{equation}\label{eq:differences_AIC}
\Delta\textrm{AIC} \equiv \textrm{AIC}_{\textrm{GR-$\Lambda$CDM}} - \textrm{AIC}_{\textrm{X}}\end{equation}
\begin{equation}
\Delta\textrm{DIC} \equiv \textrm{DIC}_{\textrm{GR-$\Lambda$CDM}} - \textrm{DIC}_{\textrm{X}}\label{eq:differences_DIC}\,.
\end{equation}
 Therefore positive differences mean that the new models ($X$) fare better than the vanilla model, whereas negative differences mean that they fare worse.
In our case, X represents either the BD-$\Lambda$CDM or the BD-RVM models. For values $0 \leq \Delta\textrm{AIC},\Delta\textrm{DIC}<2$  the level of evidence in favor of the considered option beyond the standard model is deemed as {\it weak}. If, however, $2 \leq \Delta\textrm{AIC},\Delta\textrm{DIC} < 6$ we speak of {\it positive} evidence, whereas if $6 \leq \Delta\textrm{AIC},\Delta\textrm{DIC} < 10$ it is said that there is {\it strong} evidence supporting the non-standard model. Finally, if $\Delta\textrm{AIC},\Delta\textrm{DIC}>10$  one may licitly  conclude(according to the usual argot of the information criteria) that there is {\it very strong} evidence in favor of the model under study relative to the GR-$\Lambda$CDM model. On the other hand, negative values of $\Delta\textrm{AIC},\Delta\textrm{DIC}$ would indicate that the GR-$\Lambda$CDM is the model favored by the observational data. 
\section{Discussion of the results}\label{sec:discussion}
We are now ready to discuss the results obtained in this analysis and hence to effectively assess if the considered extensions of the Brans-Dicke framework are competitive with the vanilla concordance model of cosmology (which we recall is denoted GR-$\CC$CDM in this work). The results of our analysis are collected in our tables and figures. Specifically, in
Tables \ref{tab:table_1}, \ref{tab:table_2} and \ref{tab:table_3}  we present the fitted values of the cosmological parameters for each one of the models under study, namely the vanilla model and the two Brans-Dicke extensions, i) BD-$\CC$CDM and ii) BD-RVM, which have fixed and running cosmological term, respectively. The fitting values for each model scenario are obtained by considering the three different datasets described in Sec.\,\ref{sec:Data_and_methodology}. Furthermore, in Tables \ref{tab:table_Baseline_contributions}, \ref{tab:table_Baseline_S8_contributions} and \ref{tab:table_Baseline_H0_contributions} we furnish the detailed breakdown of the different contributions to $\chi^2$, defined in Eq.\,\eqref{eq:chi2_total}, from two of the dataset scenarios  (viz. the Baseline and the Baseline+$H_0$ data sets). \jnew{Additionally, in Figs. \ref{fig:GR_LCDM_Baseline}, \ref{fig:BD_LCDM_Baseline}, \ref{fig:BD_RVM_Baseline}, \ref{fig:Baseline_all_models}, \ref{fig:GR_LCDM}, \ref{fig:BD_LCDM} and \ref{fig:BD_RVM} we provide the two-dimensional marginalized likelihood distributions for the cosmological parameters\footnote{\newnew{In particular, the purpose of Fig.\,\ref{fig:Baseline_all_models} is to ease the comparison between the considered models (presented in the previous three figures) at a glance. For example, one can see very clearly from that figure that the contour plots for the BD models favor an alleviation of the $S_8$-tension since the preferred values are smaller than in the concordance model (GR-$\Lambda$CDM). The impact on the $H_0$-tension, however, can be significant only when the Baseline+$H_0$ set is used, as seen in the subsequent figures.}} }Their mean values as well as  the mentioned Figs. \ref{fig:GR_LCDM_Baseline}-\ref{fig:BD_RVM} have been obtained with the help of the \texttt{GetDist} code \cite{Lewis:2019xzd}. Finally,  in Figs.  \ref{fig:H_z_Baseline} and \ref{fig:H_z_Baseline_H0} we confront theoretical predictions and data points for a particular source of data, as we shall comment. 

Let us now discuss our results with some more detail. As it can be seen from the caption of Table \ref{tab:table_1}, the vanilla  model has 5 free parameters $(\omega_b,n_s,A_s,H_0,\Omega^0_m)$, where we recall that  we do not consider the optical depth parameter $\tau$ since it is not sensitive to our analysis. On the other hand, the BD-$\Lambda$CDM has 6 free parameters, the new one being $\epsilon$ from \eqref{eq:power-law_BD_field}. Finally, the BD-RVM has 7 free parameters; the additional one with respect to the previous model is $\nu$ from \eqref{eq:LambdaRVM}. 
Owing to the different number of free parameters among the competing models being considered here,   we are naturally led to use the statistical information criteria AIC and DIC defined in equations \eqref{eq:AIC} and \eqref{eq:DIC}. Because these criteria penalize the extra number of parameters,  model comparison will be much fairer than just using the minimum $\chi^2$ test. For the BD-$\Lambda$CDM, and considering the Baseline dataset,  we find (cf. Table \ref{tab:table_1}) $\Delta\textrm{AIC}$ ($\Delta\textrm{DIC}$) = 4.76 (5.06),  whereas for the BD-RVM we get $\Delta\textrm{AIC}$ ($\Delta\textrm{DIC}$) = 3.81 (5.10), all of them being positive, which means, according to the definitions \eqref{eq:differences_AIC} and \eqref{eq:differences_DIC}, that in both cases there is {\it positive} evidence in favor of the non-standard models, and hence supporting the extended BD frameworks under scrutiny. Notwithstanding, it should be clear that not all the data used in our analysis show a particular sensitivity in favor of these extensions. For example, in Figs. \ref{fig:H_z_Baseline} and \ref{fig:H_z_Baseline_H0} the evolution of the Hubble function with the redshift is plotted against the observational data points and their corresponding errors. At naked eye, we can see that these data are fitted comparably well by the three models under consideration, and this is quantitatively confirmed by the corresponding $\chi^2_{H(z)}$ values in Tables \ref{tab:table_Baseline_contributions} and \ref{tab:table_Baseline_H0_contributions}. 

In point of fact, the real improvement in the fitting performance  of the models based on BD gravity with respect to the vanilla model comes from another direction and it goes hand in hand with the significantly smaller value quoted in Table \ref{tab:table_Baseline_contributions} regarding the  $\chi^2_{f\sigma_8}$ 
contribution as compared to  GR-$\CC$CDM within the Baseline dataset. This trend is confirmed with the inclusion of the $S_8$-prior  (i.e. the  Baseline+$S_8$ dataset), which increases the positive differences in the information criteria. Thus, for the BD-$\Lambda$CDM we find (cf. Table \ref{tab:table_2}) $\Delta\textrm{AIC}$ ($\Delta\textrm{DIC}$) = 7.01 (7.61) and for the BD-RVM, $\Delta\textrm{AIC}$ ($\Delta\textrm{DIC}$) = 6.39 (6.59). Both statistical criteria  point consistently towards the same conclusion, which is that when the Baseline+$S_8$ dataset is considered we find {\it strong} evidence in favor of the two BD-gravity extensions as compared with the  concordance model. This situation, however, does not replicate in exactly the same way when the alternative Baseline+$H_0$ dataset is analyzed. Here the BD model with running vacuum (the BD-RVM) stands out over the other two  and becomes the front runner. Indeed, whereas for the BD-$\Lambda$CDM we obtain the negative differences  $\Delta\textrm{AIC}$ ($\Delta\textrm{DIC}$) = -2.37 (-2.46) (cf. Table \ref{tab:table_3}),  for the BD-RVM we get, instead, the peak positive yields of our analysis:  $\Delta\textrm{AIC}$ ($\Delta\textrm{DIC}$) = 12.43 (12.10). These results illustrate, on the one hand, that the BD-$\Lambda$CDM is no better than the vanilla model (with respect to the dataset under consideration), and on the other hand show that the BD-RVM appears to be a model preferred to the vanilla model (with respect to the same dataset), this being confirmed with {\it very strong} evidence according to the jargon used by the information criteria.
On inspecting Table \ref{tab:table_Baseline_H0_contributions} we confirm once more that the origin of such an improvement lies in the significantly lower  value of $\chi^2_{f\sigma_8}$ and also, in this case, of $\chi^2_{H_0}$. In other words, its success stems  from a better description of the LSS data, and, at the same time, of the $H_0$-prior, than the vanilla model. Such an accomplishment is unmatched by the BD model with constant $\CC$, i.e. the BD-$\CC$CDM.

The obtained results are a reflex of the changes in the  values of the cosmological parameters  depending on which dataset is analyzed. Let us  focus on the change of the extra parameters with respect to the GR-$\Lambda$CDM model, namely $\epsilon$ for the BD-$\Lambda$CDM, and $\epsilon$ and $\nu$ for the BD-RVM. For the former within the Baseline dataset, we find $\epsilon = 0.0039\pm 0.0015$ (2.60$\sigma$),  whereas for the latter two we get $\epsilon = -0.0016^{+0.0037}_{-0.0049}$ (0.37$\sigma$) and $\nu=0.0019^{+0.0016}_{-0.0013}$ (1.31$\sigma$). If we now move to the Baseline+$S_8$ dataset, for the BD-$\Lambda$CDM model we get $\epsilon = 0.0042\pm 0.0015$ (2.80$\sigma$). The fact of finding a higher confidence level in favor of $\epsilon>0$ in this dataset as compared to the Baseline one should not come as a surprise since this result goes hand in hand with the increase in the positive differences of AIC and DIC observed when we move from the Baseline to the Baseline+$S_8$ dataset.

The situation with the BD framework with running vacuum (the BD-RVM model) deserves a particular consideration since the two extra parameters $\epsilon$ and $\nu$ tend to present a high level of degeneracy depending on the dataset being considered. This can be immediately appraised by looking at the panel $\epsilon$-$\nu$ in Fig. \ref{fig:BD_RVM}. There is a strong degeneracy between these two parameters when the Baseline dataset is considered and this is why we do not observe a comparable level of evidence in favor of $\epsilon\neq 0$ to that one we observe in the BD-$\Lambda$CDM model (namely $0.37\sigma$ in the former versus $2.60\sigma$ in the latter). This degeneracy, however, is broken once the $S_8$-prior enters the fit. In effect, using the Baseline+$S_8$ dataset  the BD-RVM model now renders $\epsilon = -0.0031\pm 0.0019$ (1.63$\sigma$). At the same time we find $\nu=0.00237\pm 0.00080$ (2.96$\sigma$) and hence the nonvanishing character of $\nu$ becomes now also reinforced (from $1.31\sigma$ to $2.96\sigma$). The corresponding contour plot for these parameters in Fig.\,\ref{fig:BD_RVM} also reflects this situation, as it has clearly shrunk as compared to the one of the Baseline dataset. 

Furthermore, as could be expected, the inclusion of the prior $H_0=73.04\pm 1.04$ km/s/Mpc \cite{Riess:2021jrx} has a dramatic impact on our analysis. Within the context of the BD-$\Lambda$CDM model under the Baseline+$H_0$ dataset, we find a rather inconspicuous output for the extra parameter of the model, as it is found to be compatible with zero: $\epsilon = 0.0002\pm 0.0013$ (0.15$\sigma$). As a result, the evidence in favor of a time evolving Newtonian gravitational coupling has subsided in this setup. In actual fact, the differences among the fitting values of  BD-$\Lambda$CDM and  GR-$\Lambda$CDM are now virtually absent (cf. Table \ref{tab:table_3}) and since the BD-$\CC$CDM has one additional parameter, this automatically entails a penalty from the information criteria which is reflected in the negative values of $\Delta$AIC and $\Delta$DIC in that table.  To recap, we may say that the inability of the BD-$\Lambda$CDM model to accommodate the $H_0$-prior causes the suppression of the dynamics of the BD scalar field $\phi=1/G(a)$. In stark contrast to this meager situation with fixed $\CC$ when the $H_0$-prior is included, the conclusions reached in the case with dynamical vacuum are in the opposite pole. Indeed, for the  the BD-RVM model we find (cf. Table \ref{tab:table_3}): $\epsilon = -0.00767^{+0.00069}_{-0.00220}$ (5.31$\sigma$) and $\nu=0.00332^{+0.00084}_{-0.00065}$ (4.46$\sigma$). The high confidence levels obtained for the nonvanishing values of both parameters $\epsilon$ and $\nu$ are now in the rather conspicuous range $\sim 4.5-5.3\sigma$, which is in consonance with the outstanding positive values of the information criteria in Table \ref{tab:table_3}: $\Delta\textrm{AIC}=12.43$ and  $\Delta\textrm{DIC}=12.10$  Altogether, the output of the BD-RVM for the dataset involving the $H_0$-prior brings that model to the forefront position with respect to the vanilla cosmological model and the BD-$\CC$CDM. In obtaining these results, we point out that the analysis does not show a especial sensitivity to the value and sign of the BD parameter $\omega_{\textrm{BD}}$, provided it is sufficiently large. Let us quote central values only.  For instance, in the context of the BD-$\Lambda$CDM model we find $\wBD\simeq -467$ when the Baseline dataset is considered, $\wBD\simeq -434$  for the Baseline+$S_8$ and $\wBD\simeq -9138$ for the Baseline+$H_0$. In all these cases $\epsilon>0$, implying that the effective gravitational coupling $G(a)$ increases with the expansion (i.e. for increasing $a$).
On the other hand, $\epsilon$ proves negative  for the BD-RVM within the three datasets, and hence $G(a)$ decreases with the expansion. This is favorable since the structure formation is suppressed. For this model we find, instead,  positive values of the BD parameter, to wit: $\wBD\simeq 1138$, $\wBD\simeq 587$ and  $\wBD\simeq 238$ for the Baseline,  Baseline+$S_8$ and Baseline+$H_0$ datasets, respectively.

\textcolor{black}{Let us comment on the correlations obtained between $\Omega^0_m$-$H_0$ and $H_0$-$S_8$ and the impact that they have on the tensions. In the case of the  concordance model (GR-$\CC$CDM), for the three datasets, Baseline, Baseline+$S_8$ and Baseline+$H_0$, we observe an anti-correlation between $\Omega^0_m$ and $H_0$ \jtext{and, what is more, we observe that the contour plots appear to be very narrow (see e.g. Figs. \ref{fig:GR_LCDM_Baseline} and \ref{fig:GR_LCDM}),}  which limits the allowed part of the parameter space. This was expected since this anti-correlation is at the root of the $H_0$-tension. While CMB data prefer high values of $\Omega^0_m$ and low values of $H_0$, the SH0ES data prefer exactly the opposite. As we move from the Baseline dataset ($\Omega^0_m=0.3010\pm 0.0052$) to the Baseline+$H_0$ ($\Omega^0_m=0.2942\pm 0.0048$) a reduction in the value of the matter parameter is obtained.  \jtext{However, this comes at a price since an increase  in the value of $\chi^2_{\textrm{min}}$ is obtained.} This is why the standard model of cosmology (GR-$\Lambda$CDM) cannot handle the $H_0$-tension.  \jtext{ Regarding the correlation between $H_0$ and $S_8$ within that model,  we observe from Fig. \ref{fig:GR_LCDM} that they are again negatively correlated}, as expected. \jtext{From such an anticorrelation it follows, using the relation $S_8\sim \sqrt{\Omega^0_m}$, that the higher the values of $H_0$ are the lower the values of $\Omega^0_m$ are, and hence  also the lower are those of $S_8$. Now what we expect when we deal with extra cosmological parameters is to get wider contours, and in fact it is so.  For example,  for the BD-$\Lambda$CDM, we appreciate  from Fig. \ref{fig:BD_LCDM} that} for the Baseline and Baseline+$S_8$ datasets, even if the anti-correlation between $\Omega^0_m$ and $H_0$ is still in force, \jtext{the corresponding contours are wider with respect to the GR-$\Lambda$CDM}. However, when we include the $H_0$ prior in the analysis the allowed part of the parameter-space is much more restricted. Contrary to what happens in the standard model, for the BD-$\Lambda$CDM model, $H_0$ and $S_8$ are now positively-correlated, something that can be explained due to the presence of the non-standard parameter $\epsilon$. \jtext{Finally, let us consider the situation for the BD-RVM case for which the vacuum is dynamical, see Fig. \ref{fig:BD_RVM}. Here, due to the presence of two extra cosmological parameters, $\epsilon$ and $\nu$, the anticorrelation also disappears as the contours become essentially vertical (although slightly tilted towards the corrrelated orientation between the two parameters $H_0$ and $S_8$), thus strengthening the fitted value of $H_0$ in the domain $\gtrsim 70$ km/s/Mpc. The contours are also wider and this allows for more flexibility when it comes to accommodate the observational data}. As with the other two models, we observe in such a figure that the inclusion of the $S_8$-prior and $H_0$-prior narrows down the allowed region of the parameter space, although there is still room to place the values of the parameters in the desired places. For instance, in the case where the Baseline+$H_0$ dataset is analyzed \jtext{we get a relatively large value of the Hubble parameter, specifically  $H_0 = 70.34\pm 0.63$ km/s/Mpc,}  and simultaneously a high value also for the matter parameter $\Omega^0_m = 0.2994\pm 0.0050$, \jtext{something that is impossible in the concordance model and, as we stated, it is a very welcome fact towards a possible alleviation of the tensions.}}   
                                            
It is to be emphasized that in spite of the fact that the two generalized  BD models under study can cope with the $\sigma_8$-tension for certain datasets, it turns out that only the BD-RVM model can ease the $H_0$-tension along with the $\sigma_8$-tension, and this occurs within the Baseline+$H_0$ dataset. This is manifest from the results displayed in Table \ref{tab:table_3}. Here, the vanilla (GR-$\Lambda$CDM) model yields $H_0=69.13\pm 0.38$ km/s/Mpc,  the BD-$\Lambda$CDM gives $H_0=69.02\pm 0.67$ km/s/Mpc and, finally, the BD-RVM renders the highest fitting value: $H_0= 70.34\pm 0.63$ km/s/Mpc. The residual tensions of these fitting results with respect to the latest local measurement $H_0=73.4\pm 1.04$ km/s/Mpc\,\cite{Riess:2021jrx} are 3.53$\sigma$, 3.25$\sigma$ and 2.22$\sigma$ respectively. That the BD-RVM can lower the $H_0$-tension without worsening the $\sigma_8$-tension within the Baseline+$H_0$ dataset is evinced from the fitting value of $\sigma_8$ in Table  \ref{tab:table_3}, which reads $\sigma_8 = 0.764\pm 0.018$. This is a very good result, which can be rephrased in terms of the fitting value of the related LSS observable $S_8=\sigma_8\sqrt{\Omega^0_m/0.3}$ quoted in the same table: $S_8=0.763\pm 0.019$. If we compare it with the weak-lensing measurement of this same observable from \cite{Hildebrandt:2018yau}, namely $S_8 = 0.737^{+0.040}_{-0.036}$, we find that there is no residual tension left to speak of at this point, for the mismatch is less than $1\sigma$ (specifically $\sim$0.61$\sigma$). 

Our analysis also shows that a very useful way to parametrize the background deviations of the two BD extensions with respect to the vanilla model is to consider that the BD field and the cosmological term induce an effective form of dark energy (DE) when viewed from the GR standpoint.  The deviation from $-1$ of the corresponding effective equation of state (EoS) parameter and also its dynamical character (see Figs. \ref{fig:w_eff_Baseline} and \ref{fig:example}), could be the appropriate smoking gun of the effect we are searching for.  
To be more precise, we find that the EoS signature of the BD-extension with fixed $\CC$ (the BD-$\Lambda$CDM model) is its phantom-like behavior, whereas the BD model with running $\CC$ (BD-RVM) mimics quintessence. No phantom nor quintessence fundamental fields are around, of course.

\begin{figure*}[htbp]
\centering
\mbox{\includegraphics[width=170mm]{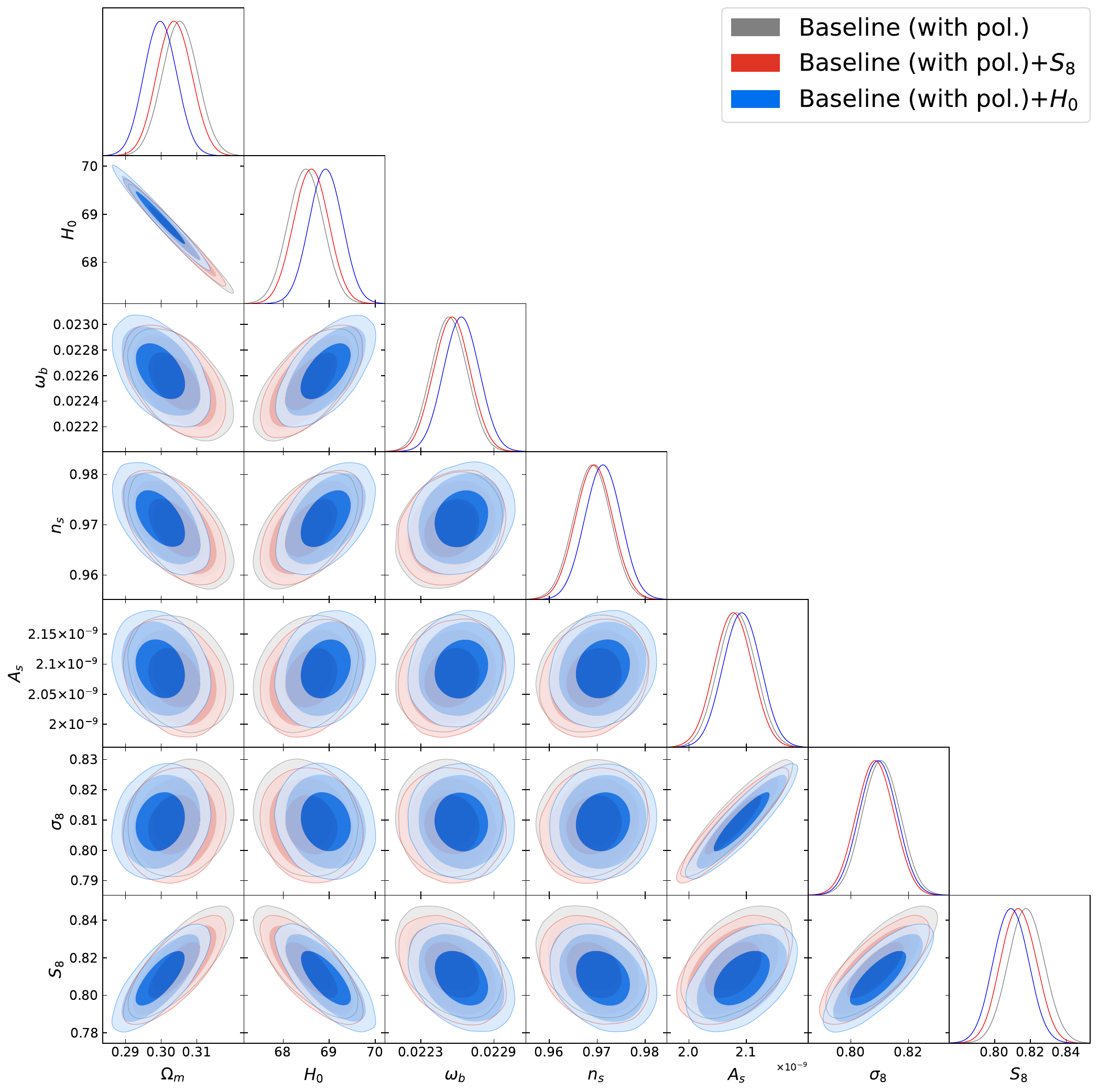}}
\caption{\textcolor{black}{The same as in Fig. \ref{fig:GR_LCDM} but for the Baseline (with pol.) (gray contours), the Baseline (with pol.)+$S_8$ (red contours) and the Baseline (with pol.)+$H_0$ (blue contours) datasets.}}
\label{fig:GR_LCDM_with_pol}
\end{figure*}

\begin{figure*}[htbp]
\centering
\mbox{\includegraphics[width=170mm]{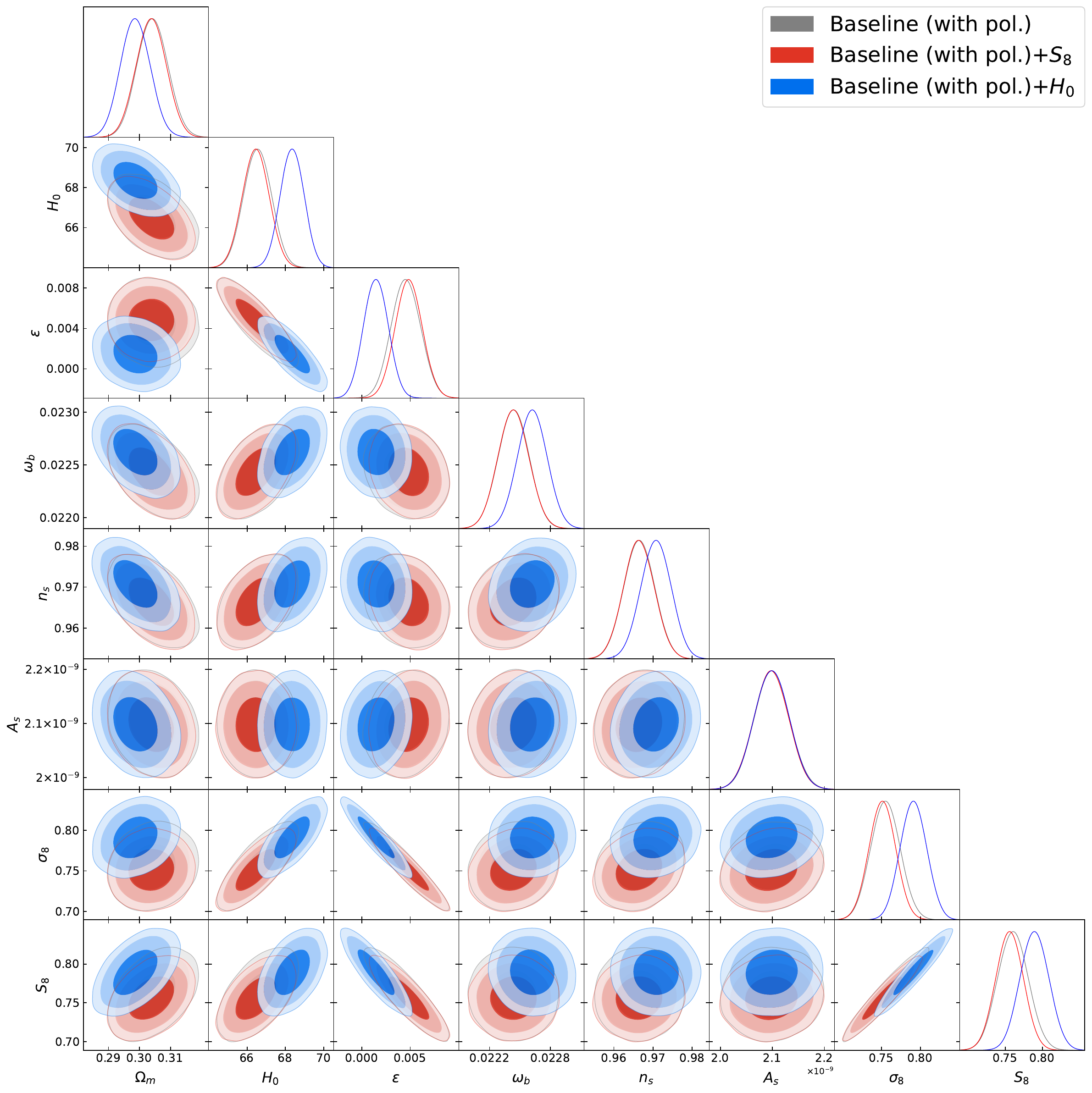}}
\caption{\textcolor{black}{Same as in Fig. \ref{fig:GR_LCDM_with_pol} but for the BD-$\Lambda$CDM cosmological model.}}
\label{fig:BD_LCDM_with_pol}
\end{figure*}

\begin{figure*}[htbp]
\centering
\mbox{\includegraphics[width=170mm]{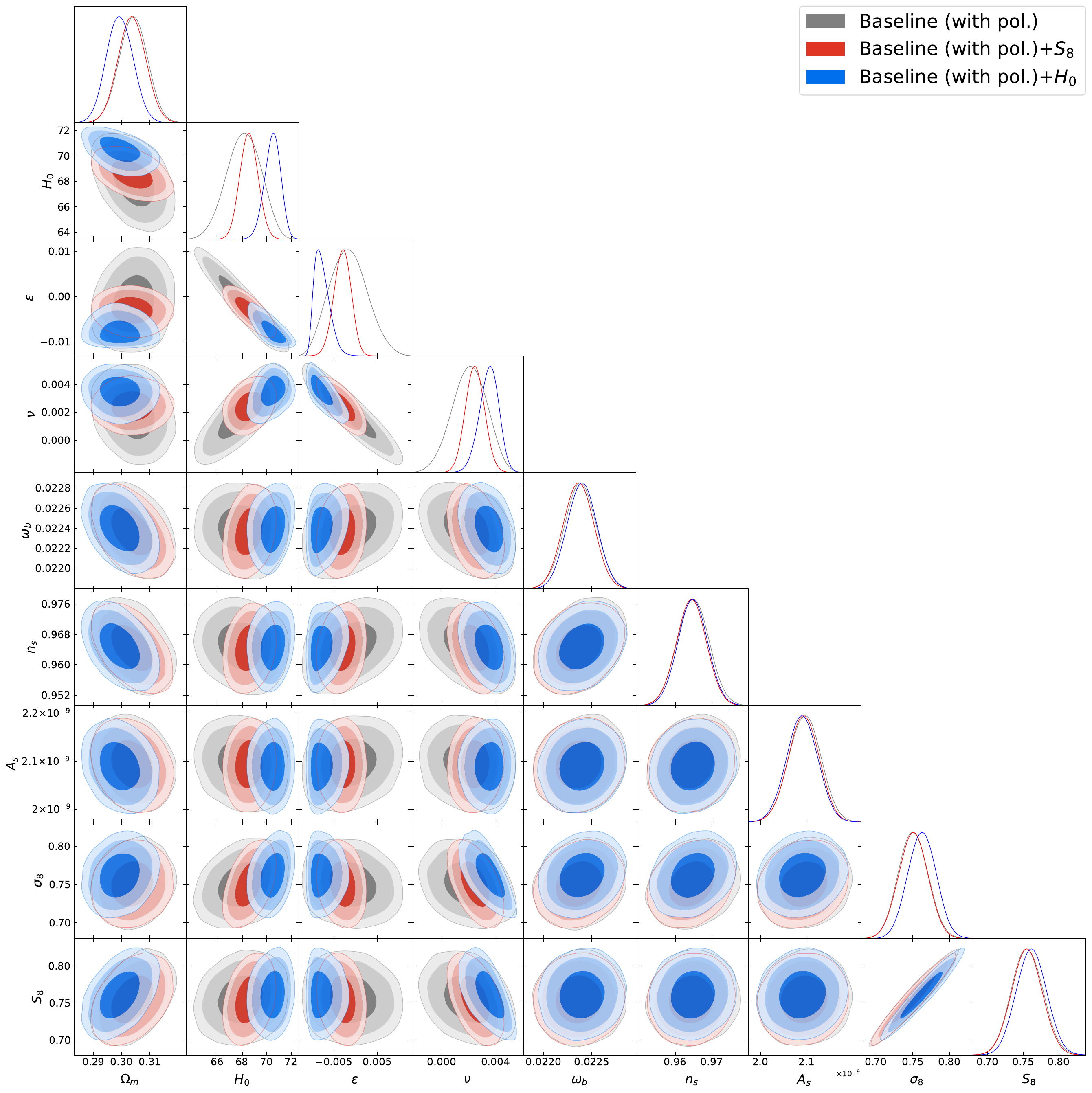}}
\caption{\textcolor{black}{Same as in Fig. \ref{fig:GR_LCDM_with_pol} but for the BD-RVM cosmological model.}
\label{fig:BD_RVM_with_pol}}
\end{figure*}

\jtext{Remarkably enough, we should mention that  when we include the polarization and the lensing data in the CMB analysis, i.e. when we use the Planck 2018 TT,TE,EE+lowE+lensing data\,\cite{Planck:2018vyg}, the main positive traits of our analysis remain in place (and as a matter of fact get improved).  The reason why we have first presented our results by treating the CMB data without polarizations is because we wanted to better compare with our previous analysis \cite{SolaPeracaula:2021gxi}, where the same kind of data are used. \jnew{At the same time, since the polarization data have been a bit controversial at some point in the literature\,\cite{Garcia-Quintero:2019cgt,Lin:2017bhs} (an issue on which we do not wish to enter here), the inclusion of the results without polarizations  makes our presentation more balanced. In the appendix we display the results of our analysis when the Planck 2018 TT,TE,EE+lowE+lensing data\,\cite{Planck:2018vyg} are employed.} The numerical results are displayed  in Tables \ref{tab:table_Baseline_with_pol}-\ref{tab:table_Baseline_with_pol_H0} and Figs. \ref{fig:GR_LCDM_with_pol}-\ref{fig:BD_RVM_with_pol}. As we can appreciate, the presence of the polarization data not only does not spoil the quality of the fit \jnew{but actually makes it better, which means that the inclusion of polarizations further magnifies the ability} of the BD-RVM in fitting the overall data and in relieving  the $H_0$ and $\sigma_8$-tensions. We can see e.g. that the corresponding values obtained for  $\sigma_8 = 0.762\pm 0.018$ and  $H_0= 70.47\pm 0.64$ km/s/Mpc within the Baseline$+H_0$ dataset are better as compared to the case without polarizations since the value of $\sigma_8$ diminishes whereas that of $H_0$ increases, see Table \ref{tab:table_Baseline_with_pol_H0} for details. Furthermore, the overall quality of the fit with CMB polarizations and lensing is remarkably higher, as can be seen from the numerical diagnostic rendered by the AIC and DIC statistical indicators (which are shown explicitly in the aforementioned  Tables \ref{tab:table_Baseline_with_pol}-\ref{tab:table_Baseline_with_pol_H0}).  The upshot is that the encouraging  conclusions that we had derived previously  from the analysis without CMB polarizations can be significantly better in the presence of polarizations. More details in the next section.}

{We end up this section  by emphasizing that the current study on extensions of the BD framework assume the FLRW metric with three-dimensional flat geometry.  However, many recent studies in the literature have exploited the properties of the non spatially flat FLRW spacetime, which can be of interest in order to tackle various conflicting aspects of the modern cosmological observations, including the $\sigma_8$ and $H_0$ tensions that are presently besetting the vanilla $\CC$CDM or concordance model of cosmology, see for instance
\cite{deCruzPerez:2022hfr,Ratra:2022ksb,Cao:2022ugh,Cao:2021ldv,Khadka:2020tlm,Cao:2020evz,Khadka:2020vlh,Park:2018bwy,Park:2018fxx,Park:2017xbl} and references therein.}

\begin{table*}
\begin{ruledtabular}
\begin{tabular}{cccc}
\\[+0.5mm]
\multicolumn{1}{c}{} & \multicolumn{3}{c}{\Large Baseline (with pol.)}
\\[+0.5mm]
\\\hline
\\[+1mm]
{ \bf Parameter} & { \bf GR-$\Lambda$CDM}  & { \bf BD-$\Lambda\textrm{CDM}$ }   & { \bf BD-$\textrm{RVM}$}\\  
\\[+1mm]
{ $\omega_b$}  & { $0.02253 (0.02253)\pm 0.00014$}  & { $0.02244 (0.02244)\pm 0.00014$}    & { $0.02238 (0.02238)\pm 0.00015$}\\
{$n_s$}   & {$0.9690 (0.9690)\pm 0.0036$} & {$0.9665 (0.9665)\pm 0.0036$} & {$0.9649 (0.9648)\pm 0.0040$}\\
{$10^{9}A_s$}    & {$2.083 (2.083)\pm 0.030$} & {$2.099 (2.099)\pm 0.031$}  & { $2.095 (2.095)\pm 0.032$}\\
{$H_0$[km/s/Mpc]}   & {$68.49 (68.49)\pm 0.35$} & { $66.55 (66.54)\pm 0.67$}  & { $68.1 (68.24)^{+1.5}_{-1.3}$}\\  
{$\Omega^0_m$}   & { $0.3053 (0.3052)\pm 0.0046$} & { $0.3042 (0.3042)\pm 0.0046$}   & { $0.3041 (0.3039)\pm 0.0047$}\\  
{$\sigma_8(0)$}   & {$0.8105 (0.8105)\pm 0.0060$} & { $0.755 (0.755)\pm 0.017$}     & { $0.750 (0.754)\pm 0.019$}\\   
{$S_8$}   & {$0.8177 (0.8175)\pm 0.0093$} & { $0.761 (0.760)\pm 0.019$}     & { $0.756 (0.759)\pm 0.020$}\\   
{$\epsilon$}   & {-} & { $0.0046 (0.0046)\pm 0.0014$}     & { $-0.0016 (-0.0020)^{+0.0040}_{-0.0046}$}\\   
{$\nu$}   & {-} & {-}    & {$0.0020 (0.0021)\pm 0.0012$}\\   
{$\chi^2_{\rm min}$}   & { 76.86} & { 64.31}    & { 62.37}\\  
{$\Delta$AIC}   & { -} & { 9.98}     & { 9.43 }\\
{$\Delta$DIC}   & { -} & { 10.65}     & { 10.03 }\\
\end{tabular}
\\[+1mm]
\end{ruledtabular}
\caption{\textcolor{black}{As in Table \ref{tab:table_1} but replacing the Planck 2018 TT+lowE (see \eqref{eq:correlationCMB}) data with the Planck TT,TE,EE+lowE+lensing (see \eqref{eq:correlationCMB_with_pol}).  }} 
\label{tab:table_Baseline_with_pol}
\end{table*}
\begin{table*}
\begin{ruledtabular}
\begin{tabular}{cccc}
\\[+0.5mm]
\multicolumn{1}{c}{} & \multicolumn{3}{c}{\Large Baseline (with pol.)+$S_8$}
\\[+0.5mm]
\\\hline
\\[+1mm]
{ \bf Parameter} & { \bf GR-$\Lambda$CDM}  & { \bf BD-$\Lambda\textrm{CDM}$ }   & { \bf BD-$\textrm{RVM}$}\\  
\\[+1mm]
{ $\omega_b$}  & { $0.02256 (0.02255)\pm 0.00014$}  & { $0.02244 (0.02244)\pm 0.00014$}    & { $0.02236 (0.02237)\pm 0.00015$}\\
{$n_s$}   & {$0.9694 (0.9695)\pm 0.0036$} & {$0.9664 (0.9664)\pm 0.0036$} & {$0.9644 (0.96452)\pm 0.0038$}\\
{$10^{9}A_s$}    & {$2.077 (2.081)\pm 0.030$} & {$2.098 (2.098)\pm 0.031$}  & { $2.093 (2.094)\pm 0.031$}\\
{$H_0$[km/s/Mpc]}   & {$68.61 (68.61)\pm 0.35$} & { $66.46 (66.45)\pm 0.65$}  & { $68.45 (68.51)\pm 0.68$}\\  
{$\Omega^0_m$}   & { $0.3037 (0.3036)\pm 0.0045$} & { $0.3038 (0.3038)\pm 0.0045$}   & { $0.3036 (0.3036)\pm 0.0045$}\\  
{$\sigma_8(0)$}   & {$0.8084 (0.8084)\pm 0.0060$} & { $0.751 (0.751)\pm 0.016$}     & { $0.756 (0.736)\pm 0.017$}\\   
{$S_8$}   & {$0.8134 (0.8133)\pm 0.0091$} & { $0.756 (0.756)\pm 0.017$}     & { $0.751 (0.740)\pm 0.018$}\\   
{$\epsilon$}   & {-} & { $0.0048 (0.0049)\pm 0.0013$}     & { $-0.0031 (-0.0029) \pm 0.0018$}\\   
{$\nu$}   & {-} & {-}    & {$0.00245 (0.00239)\pm 0.00065$}\\   
{$\chi^2_{\rm min}$}   & { 81.11} & { 64.61}    & { 62.42}\\  
{$\Delta$AIC}   & { -} & {14.11}     & { 13.82 }\\
{$\Delta$DIC}   & { -} & { 14.35 }     & { 14.15 }\\
\end{tabular}
\\[+1mm]
\end{ruledtabular}
\caption{\textcolor{black}{As in Table \ref{tab:table_Baseline_with_pol} but adding as an input the $S_8\equiv\sigma_8\sqrt{\Omega^0_m/0.3} = 0.737^{+0.040}_{-0.036}$ weak-lensing measurement from \cite{Hildebrandt:2018yau}.}}
\end{table*}
\begin{table*}
\begin{ruledtabular}
\begin{tabular}{cccc}
\\[+0.5mm]
\multicolumn{1}{c}{} & \multicolumn{3}{c}{\Large Baseline (with pol.)+$H_0$}
\\[+0.5mm]
\\\hline
\\[+1mm]
{ \bf Parameter} & { \bf GR-$\Lambda$CDM}  & { \bf BD-$\Lambda\textrm{CDM}$ }   & { \bf BD-$\textrm{RVM}$}\\  
\\[+1mm]
{ $\omega_b$}  & { $0.02263 (0.02263)\pm 0.00014$}  & { $0.02262 (0.02262)\pm 0.00014$}    & { $0.02240 (0.02233)\pm 0.00014$}\\
{$n_s$}   & {$0.9712 (0.9713)\pm 0.0035$} & {$0.9707 (0.9708)\pm 0.0036$} & {$0.9648 (0.9629)\pm 0.0037$}\\
{$10^{9}A_s$}    & {$2.092 (2.092)\pm 0.030$} & {$2.098 (2.099)\pm 0.031$}  & { $2.089 (2.084)\pm 0.031$}\\
{$H_0$[km/s/Mpc]}   & {$68.92 (68.93)\pm 0.34$} & { $68.35 (68.36)\pm 0.57$}  & { $70.47 (71.32)^{+0.70}_{-0.59}$}\\  
{$\Omega^0_m$}   & { $0.2998 (0.2997)\pm 0.0042$} & { $0.2987 (0.2987)\pm 0.0044$}   & { $0.2993 (0.3000)\pm 0.0043$}\\  
{$\sigma_8(0)$}   & {$0.8094 (0.8094)\pm 0.0060$} & { $0.791 (0.791)\pm 0.016$}     & { $0.762 (0.643)\pm 0.018$}\\   
{$S_8$}   & {$0.8091 (0.8090)\pm 0.0090$} & { $0.789 (0.790)\pm 0.018$}     & { $0.761 (0.642)\pm 0.019$}\\   
{$\epsilon$}   & {-} & { $0.0015 (0.0014)\pm 0.0012$}     & { $-0.0077 (-0.0102)^{+0.0012}_{-0.0021}$}\\   
{$\nu$}   & {-} & {-}    & {$0.00348 (0.0442)^{+0.00077}_{-0.00064}$}\\   
{$\chi^2_{\rm min}$}   & { 94.15} & { 92.50}    & { 70.04}\\  
{$\Delta$AIC}   & { -} & { -0.74}     & { 19.24 }\\
{$\Delta$DIC}   & { -} & { -0.28}     & { 19.56 }\\
\end{tabular}
\\[+1mm]
\end{ruledtabular}
\caption{\textcolor{black}{The same as in Table \ref{tab:table_Baseline_with_pol} but considering the prior on $H_0=73.04\pm 1.04$ km/s/Mpc from \cite{Riess:2021jrx}.} }
\label{tab:table_Baseline_with_pol_H0}
\end{table*}
%
%



%
%
\section{Conclusions}\label{sec:conclusions}
In this work we have found intriguing signs that a mild (logarithmic) time-variation of the Newtonian coupling in combination with a slow cosmic evolution of the vacuum energy density (VED) can be essential ingredients to improve the global quality fit to the modern cosmological data and relieve the $\CC$CDM tensions. The framework that we have used to realize this scenario is the time-honored Brans \& Dicke (BD) gravitational theory\,\cite{BransDicke1961}, where the time variation of the Newtonian coupling is implemented through the slowly varying BD field $\phi$ coupled to curvature.  In contrast to the original BD theory, however, we have admitted the presence of a cosmological term, which we have treated here either as a rigid VED $\rv=$ const. or as a `running' one, namely as a function  $\rv(H)$ evolving quadratically with the Hubble rate $H$, \jnew{based on recent results from QFT in curved spacetime pointing to such an evolution, cf. \cite{Moreno-Pulido:2022phq,Moreno-Pulido:2020anb,Moreno-Pulido:2022upl,Moreno-Pulido:2023ryo}.} The rigid VED option leads to what we have called the BD-$\CC$CDM model, which is the BD counterpart of the usual concordance $\CC$CDM model in GR (denoted here GR-$\CC$CDM  for obvious reasons). \jtext{As for the dynamical option, we have chosen the very specific form for $\rv(H)$ which is predicted within the context of the running  vacuum model (RVM), where the change of the VED with the cosmic expansion goes as  $\delta\rho_{\textrm{vac}}(H)\propto \nu\, m^2_{\textrm{Pl}} (H^2-H_0^2)$ (with $|\nu|\ll 1$}). This expression has been recently derived from first principles, namely within QFT in FLRW spacetime, see the above mentioned references and also the reviews \cite{SolaPeracaula:2022hpd,SolaPeracaula:2023wqw}. Such a mild form of vacuum dynamics, in combination with the BD gravity framework, leads to what we have called the BD-RVM model,  in which $\rv$ changes quadratically with $H$ through a small coefficient $|\nu|\sim 10^{-3}$ associated with the $\beta$-function of the running VED in QFT in cosmological spacetime. That coefficient is precisely computable (analytically)  from the quantum effects associated with the effective action of gravity with quantized matter fields, as shown in the aforementioned references. However, its ultimate numerical value must be determined by comparison with the observational data.

\jnew{The results that we have obtained for the BD-$\Lambda$CDM and  BD-RVM models show, according to the information criteria, that for the Baseline dataset (SNIa+$H(z)$+BAO+LSS+CMB) -- which should be understood as our main dataset -- it is possible to improve the performance of the standard model of cosmology. To be specific, the results read as follows:}

\jnew{i) Without CMB polarizations: For the BD-$\Lambda$CDM we obtain $\Delta$AIC ($\Delta$DIC) = +4.76 (+5.06), whereas for the BD-RVM we get $\Delta$AIC ($\Delta$DIC) = +3.81 (+5.10), which means that from the point of view of the information criteria both models are {\it positively} favored over the GR-$\Lambda$CDM model;}

\jnew{ii) With CMB polarizations: For the BD-$\Lambda$CDM we find $\Delta$AIC ($\Delta$DIC) = +9.98 (+10.65), whereas for the BD-RVM we obtain $\Delta$AIC ($\Delta$DIC) = +9.43 (+10.03). Clearly, the presence of polarizations increases substantially the overall quality of the fit in both cases and then both models become {\it strongly} favored over the GR-$\Lambda$CDM model, if we  adopt the jargon of the information criteria.}

\jnew{We should also stress that among the three different cosmological models that we have studied, only the  BD-RVM (i.e. the one with mild dynamical vacuum energy density and also a small variation of the gravitational constant) has the capability to reduce the $H_0$-tension (without worsening the $S_8$-tension). For example, the value of the Hubble parameter that we have obtained for that model within the  Baseline dataset without polarizations is $H_0 = 68.0^{+1.4}_{-1.2}$ km/s/Mpc, hence only 3$\sigma$ away from the SH0ES measurement. The corresponding value for the case with CMB polarizations is very similar.  As indicated, this is achieved without impairing the $S_8$-tension, and actually reducing it significantly. In both cases the value of $S_8$ is in the ballpark of $0.75-0.77$ and hence in the low range. This notable improvement (i.e. lessening) of the $S_8$-tension also occurs for the BD-$\Lambda$CDM, but it is not accompanied with an amelioration of the $H_0$-tension. Nonetheless, the overall result of our fitting analysis of the Baseline data is that the two type of extended BD models (with rigid and variable vacuum energy density) are {\it positively} to {\it strongly} favored by the information criteria over the concordance model, as explicitly quantified in our fitting tables. }


\jnew{That said and established, as far as the Baseline data fits are concerned, we have also found that the BD-RVM model could be further enhanced if we include either the $S_8$-prior or the $H_0$-prior in the analysis. However, as noted in Section \ref{sec:Data_and_methodology}, the potential improvement gained in these cases is to be interpreted with care, as priors may bias the models' performance towards the desired direction. Whether it is so, however, will depend on the inherent models' ability to contribute to fixing the tensions. Thus, we emphasize it again,  it is only indicative, but it still can be useful as it constitutes a means to unveil the potential capability of the cosmological models in relieving the tensions. Moreover, the biased dragging is not automatic, as can be checked e.g. in  the case of the BD model with rigid cosmological constant (i.e. the BD-$\CC$CDM). Indeed, from the output of the information criteria shown in Tables II and III, or VIII and IX, we can confirm that this model provides a better fit than the concordance model in the presence of the $S_8$-prior, whilst, in contrast, it yields a worse fit when we employ the $H_0$-prior. In short, the mentioned improvements with respect to the GR-$\Lambda$CDM model should be interpreted only as positive hints suggesting that a given model may alleviate the $H_0$ and/or the $\sigma_8$ tensions.  In the  particular case of the BD-RVM, the use of either prior (separately) provides  intriguing positive hints towards alleviating both tensions at a time (cf. Tables II-III for the case without CMB polarizations, and VIII-IX when the CMB polarizations and lensing are included). A more detailed analysis will be needed to confirm these results. }




\jtext{To summarize, in the absence of priors (i.e. the Baseline situation), and irrespective of using or not using the CMB polarization data, the two BD-models (BD-$\CC$CDM and BD-RVM) perform better than the concordance model (GR-$\CC$CDM), if we attend to the quantitative output of the information criteria (cf. Tables I and VIII). This is our main conclusion in this analysis. Let us remark that with the inclusion of the polarization data the yield of the two BD models under Baseline data is outstanding since both information criteria, AIC and DIC, provide $\Delta$AIC, $\Delta$DIC$>+9$ with respect to GR-$\CC$CDM, which in the jargon of these criteria implies \textit{strong} evidence in favor of the two BD type of models versus the concordance model. Finally, in the presence of one of the priors ($S_8$ or $H_0$), the BD model with rigid vacuum energy (BD-$\CC$CDM) reacts positively to $S_8$ but negatively to $H_0$, whereas the the BD model with running vacuum energy (BD-RVM) reacts very positively to the presence of either prior separately, and this holds both with or without polarizations (see Tables III and IX),  reaching \textit{very strong evidence} in favor of this model: $\Delta$AIC, $\Delta$DIC$>+12$. This may be indicative of the ability of that model to relieve the two tensions at the same time. Let us mention that when we include the $H_0$-prior, the BD-RVM model yields fitting values for the parameters which attain  very high confidence levels:  $\epsilon = -0.00767^{+0.00069}_{-0.00220}$ (5.31$\sigma$) and $\nu=0.00332^{+0.00084}_{-0.00065}$ (4.46$\sigma$). These results suggest that the extra (i.e. beyond the standard model) contributions induced by these parameters are definitely welcome; in other words, they are highly preferred to be non-vanishing rather than vanishing, in full consistency with  the large values recorded by the information criteria.  The fact that the BD-RVM performance  can be greatly enhanced with the presence of either prior ($S_8$ or $H_0$) might be indicative of the  superior capability of that model to fit the data and to cope with the two tensions. 
In future work, however, we expect to be able to fully confirm these promising results in the context of a more detailed study.}
\begin{acknowledgments}
JSP is funded by projects  PID2019-105614GB-C21 (MINECO), 2021-SGR-249 (Generalitat de Catalunya) and CEX2019-000918-M (ICCUB).    He also  acknowledges  the participation in  the COST Association Action CA18108  ``Quantum Gravity Phenomenology in the Multimessenger Approach  (QG-MM)''. JSP and JdCP  acknowledge as well their participation in the COST Association Action CA21136 ``Addressing observational tensions in cosmology
with systematics and fundamental physics'' (CosmoVerse). JdCP was supported by a FPI fellowship associated to the project FPA2016-76005-C2-1-P (MINECO). 

\end{acknowledgments}


\appendix

\section{Results with Planck 2018 TT,TE,EE+lowE+lensing data}\label{AppendixA}
\textcolor{black}{For completeness, in this appendix we present the numerical results when the Planck 2018 TT+lowE data described in detail in Sec. \ref{sec:Data_and_methodology} are replaced with the Planck 2018 TT,TE,EE+lowE+lensing data\,\cite{Planck:2018vyg}. The mean values and the corresponding error bars for the different components are $\omega_b = 0.02239\pm 0.00015$, $A_s = (2.102\pm 0.031)\times 10^{-9}$, $n_s = 0.9657\pm 0.0041$, $\ell_a=301.529\pm 0.083$ and  $\mathcal{R}=1.7497\pm 0.0040$. We also provide the matrix containing the correlations between the different elements:
\begin{small}
\begin{equation}\label{eq:correlationCMB_with_pol}
C_{\textrm{CMB}} = 
\begin{pmatrix}
1 & 0.2297 & 0.4210 & -0.2164 & -0.6417  \\
0.2297 & 1 & 0.2713 & -0.1303 & -0.3063  \\
0.4210 & 0.2713 & 1 & -0.2590 & -0.7066   \\
-0.2164 & -0.1303 & -0.2590 & 1 & 0.3553  \\
-0.6417 & -0.3063 & -0.7066 & 0.3553 & 1 .
\end{pmatrix}
\end{equation}
\end{small} 
As it is clear from Tables \ref{tab:table_Baseline_with_pol}-\ref{tab:table_Baseline_with_pol_H0}, the inclusion of the polarization data enhances the performance of the BD-$\Lambda$CDM and the BD-RVM models as compared with the GR-$\Lambda$CDM. For example, for the Baseline (with pol.) case the values of the $\Delta$AIC and the $\Delta$DIC have increased with respect to the Baseline case, now indicating that both models are on the verge of being {\it very strongly} favoured over the GR-$\Lambda$CDM model. For the Baseline (with pol.)+$S_8$ the values of the information criteria (in favor of the models under consideration) are even larger since for both models we find $\Delta$AIC and $\Delta$DIC $\sim$ 14, which again means that they are {\it very strongly} favoured (in the usual parlance of the information criteria) with respect to the standard model. Finally, when the Baseline (with pol.)+$H_0$ dataset is analyzed, we observe that the BD-$\Lambda$CDM model is no longer favoured over the GR-$\Lambda$CDM. Only the BD-RVM model can (simultaneously) accommodate large values of the $H_0$ parameter ($70.47^{+0.70}_{-0.59}$ km/s/Mpc) and low values of the $\sigma_8$ parameter ($0.762\pm 0.018$). As for the fitting values of the extra parameters,  we find $\epsilon = -0.0077^{+0.0012}_{-0.0021}$ (4.67$\sigma$ c.l.) and $\nu = 0.00348^{+0.00077}_{-0.00064}$ (4.94$\sigma$ c.l.). The level of evidence for a nonvanishing $\nu$ value, namely the parameter is responsible for the running of the vacuum energy density, becomes remarkable.  }

\newpage

\bibliographystyle{ieeetr}
\bibliography{references}

\end{document}